\documentclass[12pt]{article}
\usepackage{amsmath,amssymb,amsfonts,amsthm,bbm}
\usepackage{epic,eepic,epsfig,longtable}
\usepackage{multirow,verbatim}
\usepackage{array}
\usepackage{graphicx}
\usepackage{floatrow}
\usepackage{subcaption}
\usepackage{paralist}
\usepackage{latexsym}
\usepackage{comment}
\usepackage{epsfig}
\usepackage{setspace}
\usepackage{CJK}
\usepackage{color}

\usepackage{geometry}
\usepackage{algorithm}
\usepackage{algpseudocode}

\usepackage{rotating}

\usepackage[authoryear, round]{natbib}
\usepackage{xcite}
\usepackage{xr}
\makeatletter
\newcommand*{\addFileDependency}[1]{
  \typeout{(#1)}
  \@addtofilelist{#1}
  \IfFileExists{#1}{}{\typeout{No file #1.}}
}
\makeatother
\newcommand*{\myexternaldocument}[1]{%
    \externaldocument{#1}%
    \addFileDependency{#1.tex}%
    \addFileDependency{#1.aux}%
}
\myexternaldocument{OGI_Supplement}

\def\spacingset#1{\renewcommand{\baselinestretch}%
{#1}\small\normalsize} \spacingset{1}

\makeatletter
\def\singlespace{\def\baselinestretch{1}\@normalsize}

\numberwithin{equation}{section}

\renewcommand{\hat}{\widehat}

\renewcommand{\hat}{\widehat}

\newcommand{\bfm}[1]{\ensuremath{\mathbf{#1}}}

   \def\bA{\bfm A}

    \def\FF{\mathbb{F}}

\newcommand{\bfsym}[1]{\ensuremath{\boldsymbol{#1}}}

\def\1{\bfsym{1}}	

\DeclareMathOperator{\E}{E}

\DeclareMathOperator{\Var}{Var}
\DeclareMathOperator{\var}{var}

\def\var{\mbox{var}}

\def\today{\ifcase\month\or
  January\or February\or March\or April\or May\or June\or
  July\or August\or September\or October\or November\or December\fi
  \space\number\day, \number\year}

\newdimen\biblioindent    \biblioindent=30pt

 at 8truept

\newcommand{\beq}{\begin{equation}}
  \newcommand{\eeq}{\end{equation}}
\newcommand{\beqn}{\begin{eqnarray}}
  \newcommand{\eeqn}{\end{eqnarray}}
\newcommand{\beqnn}{\begin{eqnarray*}}
  \newcommand{\eeqnn}{\end{eqnarray*}}

\allowdisplaybreaks
\usepackage{verbatim}
\def\FF{\mathcal{F}}
\def\[{\left [}  \def\]{\right ]} \def\({\left (}  \def\){\right )}
 
\def\hat{\widehat}

\newtheorem{assumption}{Assumption}
\newtheorem{theorem}{Theorem}

\newtheorem{corollary}{Corollary}

\theoremstyle{definition}
\newtheorem{definition}{Definition}

\newtheorem{remark}{Remark}

\title{Overnight  GARCH-It\^o Volatility Models }
\author{Donggyu Kim$^a$, Minseok Shin$^a$, and Yazhen Wang$^b$ \\
$^a$College of Business, \\
Korea Advanced Institute of Science and Technology (KAIST), \\
$^b$Department of Statistics, University of Wisconsin-Madison
}

\begin{document}
\maketitle
\begin{spacing}{1.9}

\begin{abstract}
Various parametric volatility models for financial data have been developed to incorporate high-frequency realized volatilities 
and better capture market dynamics. However, because high-frequency trading data are not available during the close-to-open period, the volatility models  often ignore volatility information over the close-to-open period and thus may suffer from 
loss of important information relevant to 
market dynamics. In this paper, to account for whole-day market dynamics, we propose an overnight volatility model based on It\^o diffusions to accommodate two different  instantaneous volatility processes for the open-to-close and close-to-open periods.   
We develop a weighted least squares method to estimate model parameters for two different periods and investigate its asymptotic properties. 
We conduct a simulation study to check the finite sample performance of the proposed model and method. Finally, we apply 
 the proposed approaches to real trading data. 
\end{abstract}



\section{Introduction} \label{SEC-1}
 
 Since \citet{markowitz1952portfolio} introduced the modern portfolio theory, measuring risk has become important in financial applications. Volatility itself  is often employed as a proxy for risk. Furthermore, there are several risk measurements, such as Value at Risk (VaR), expected shortfall, and market beta \citep{duffie1997overview, rockafellar2000optimization, sharpe1964capital}. These risk measurements take volatilities as an important ingredient in their formulations, and their performances heavily depend on the accuracy of volatility estimation.  
 
Generalized autoregressive conditional heteroskedasticity (GARCH) models are one of the most successful volatility models 
for low-frequency data \citep{bollerslev1986generalized, engle1982autoregressive}.
They employ squared daily log-returns as innovations in conditional expected volatilities  and are able to capture low-frequency market dynamics, such as volatility clustering and heavy tail. 
At the high-frequency level, nonparametric approaches, such as It\^o processes and realized volatility estimators, are 
often utilized to model and estimate volatilities. 
Examples include  two-time scale realized volatility (TSRV) \citep{zhang2005tale}, multi-scale realized volatility (MSRV) \citep{zhang2006efficient}, kernel realized volatility (KRV) \citep{barndorff2008designing},
quasi-maximum likelihood estimator (QMLE) \citep{ait2010high, xiu2010quasi}, pre-averaging realized volatility (PRV) \citep{jacod2009microstructure}, and robust pre-averaging realized volatility \citep{fan2018robust}.
In practice, we often observe jumps in financial data, and the decomposition of daily variation into continuous and jump components can improve 
volatility estimation and aid with better explanation of volatility dynamics \citep{ait2012testing,   barndorff2006econometrics, corsi2010threshold}.
For example, \citet{fan2007multi} and \citet{zhang2016jump} employed the wavelet method to identify the jumps in given noisy high-frequency data. \citet{mancini2004estimation} studied a threshold method for jump-detection and presented the order of an optimal threshold, and \citet{jacod2009microstructure} introduced the jump robust pre-averaging realized (PRV) estimator.
We call this realized volatility. 
There have been several recent attempts to combine low-frequency GARCH and SV models and high-frequency realized volatilities. 
Examples include the realized volatility based modeling approaches 
\citep{ andersen2003modeling}, 
the heterogeneous auto-regressive (HAR) models \citep{corsi2009simple},  the high-frequency based volatility (HEAVY) models \citep{shephard2010realising}, the realized GARCH models \citep{hansen2012realized}, and the unified GARCH/SV-It\^o models 
\citep{kim2019factor, kim2016unified,  Song2018Volatlity}. The realized volatility based models, such as HAR, HEAVY, and realized GARCH models, take reduced ARFIMA forms to model and forecast realized volatilities estimated from high-frequency data, and the unified GARCH/SV-It\^{o} models provide theoretical 
platform to reconcile low-frequency GARCH/SV volatility representations and high-frequency volatility processes and harnesses realized volatilities 
and GARCH/SV models to yield better, albeit more complicated, modeling and inference for combining low- and high-frequency data.  
Empirical studies have shown that, with realized volatility as a part of the innovation, volatility models can better capture market dynamics. 
However, because high-frequency data are usually available only during trading hours, such as the open-to-close period, the high-frequency volatility models often include open-to-close integrated volatility in the innovation and ignore the overnight risk \citep{corsi2009simple, kim2016unified,  Song2018Volatlity}.  
 \citet{taylor2007note} showed that the overnight information is important for  evaluating risk management models, so the volatility measured by the open-to-close high-frequency observations may significantly undervalue their risk.
Furthermore, the overnight risk  is often severe---for example, during the European debt crisis, Asian financial crisis, and so on---and so it is an important factor that accounts for market dynamics. From this point of view, there are several studies on  the impact of overnight returns on volatility and modeling the volatility process using overnight returns and realized volatility. 
 \citet{hansen2005realized} studied optimal incorporation of the overnight information and proposed inverse weighting of the realized
volatility and squared overnight returns by using the corresponding variance estimates.
\citet{andersen2011reduced} modeled the overnight returns using an augmented GARCH type structure.
See also \citet{martens2002measuring, todorova2014overnight, tseng2012impact} for more information on the impact of overnight volatility. 
These studies document an increasing interest in developing It\^o process-based models that provide a rigorous mathematical 
formulation for using both open-to-close high-frequency data and close-to-open low-frequency data to analyze whole-day market dynamics. 

 In this paper, we develop an instantaneous volatility model for a whole-day period.
 The whole-day is broken down into two time periods, the open-to-close and close-to-open periods.
During the open-to-close period, we observe high-frequency trading data, whereas during the close-to-open period, we observe  low-frequency close and open prices.
To reflect this structural difference, we develop two different instantaneous volatility processes for the open-to-close and close-to-open periods. 
For example, for the open-to-close period, we use the current integrated volatility as an innovation to reflect the market dynamics immediately, which helps to adapt to the rapid change in the volatility process, as occurs in the high-frequency volatility models \citep{corsi2009simple, hansen2012realized, shephard2010realising, Song2018Volatlity}. 
For the close-to-open period, we employ the current squared log-return as an innovation, which brings us back to the discrete-time GARCH model for the close-to-open period. 
The proposed structure implies that the conditional expected volatility for the whole-day period is a function of past open-to-close integrated volatilities  and squared close-to-open log-returns. 
 We call this volatility model the overnight GARCH-It\^o (OGI) model. Moreover, to estimate its model parameters, we develop a quasi-likelihood estimation procedure.
 Specifically, for the open-to-close period, we employ realized volatilities as a proxy for the corresponding conditional expected volatilities, whereas for the close-to-open period, we adopt squared close-to-open log-returns as a proxy for the corresponding conditional expected volatilities.  
These proxies have heterogeneous variances that are related to the accuracy of the proxies.
To reflect this, we calculate their variances and assign different weights to each proxy. As a result, the proposed estimation method takes the form of weighted least squares.
We apply the overnight GARCH-It\^o model for a VaR study. 

 The rest of this paper is organized as follows. 
 Section \ref{SEC-2} introduces the overnight GARCH-It\^o model and discusses its properties. 
Section \ref{SEC-WLSE} proposes weighted least squares estimation methods and investigates its asymptotic properties. 
Section \ref{SEC-simulation} conducts a simulation study to check the finite sample performance of the proposed estimation methods. Section \ref{SEC-empirical} applies the proposed overnight GARCH-It\^o model and method to real trading data. 
The conclusion is presented in  Section \ref{SEC-6}.
We collect the proofs in the Supplement document.


\setcounter{equation}{0}
\section{Overnight GARCH-It\^o models} \label{SEC-2}

In this section, we develop an It\^o diffusion process to capture the whole-day market dynamics.
To separate the  parameters for the high-frequency period (open-to-close) and low-frequency period (close-to-open), we use the subscript or superscript $H$ and $L$, respectively. 
For the low-frequency GARCH volatility related parameter, we use superscript $g$. 

\begin{definition}\label{Def-1}
We call the log-price $X_t$ an overnight GARCH-It\^o (OGI) process if it satisfies
\begin{eqnarray*}
	&&dX_t= \mu_t dt + \sigma_t (\theta) dB_t + J_t  d \Lambda_t  , \cr
	&&\sigma_{t}^2(\theta)   =
	 \begin{cases}
 \sigma^2 _{[t]}(\theta) +  \frac{ (t- [t])^2}{ \lambda ^2}( \omega_{H1}+   \gamma_H  \sigma_{[t]}^2(\theta)) -  \frac{t- [t] }{ \lambda }   ( \omega_{H2}+ \sigma_{[t]}^2(\theta))  \\
+   \frac{  \beta _H(t-[t]) ([t] +\lambda-t)  }{  \lambda ^2 (1-\lambda)}  \sum_{j=1}^\infty \gamma ^{j-1}   \( \int_{[t]+\lambda-j } ^{[t]+1-j} \sigma_s  (\theta) dB_s  \)^2 \\
 + \frac{ \alpha _H}{ \lambda } \int_{[t]} ^{t} \sigma_s ^2 (\theta) ds  + \frac{ \nu_H}{ \lambda^{2}} \left( [t]+ \lambda -t \right) (Z _t ^H)^2, & \text{ if } t \in  \( [t],  [t] +\lambda \], \\ 
\sigma^2 _{[t]+\lambda}(\theta) + \frac{  t- [t]-\lambda  }{ 1-\lambda} ( \omega_{L}+  (\gamma_L -1) \sigma_{[t]+\lambda}^2(\theta)) \\
	 +\frac{ \alpha _L(t- [t]-\lambda) ([t]+1-t) }{(1-\lambda)^2 \lambda }   \sum_{j=1}^\infty \gamma ^{j-1}  \int_{[t]+1-j} ^{[t]+1 -j +\lambda} \sigma_s ^2 (\theta) ds  \\
	 +  \frac{\beta _L}{  1-\lambda}\( \int_{[t]+\lambda} ^{t} \sigma_s  (\theta) dB_s  \)^2	  + \frac{\nu_L}{(1- \lambda)^{2}} \left( [t]+ 1 -t \right) (Z _t ^L)^2,& \text{ if } t \in  \( [t] +\lambda,  [t] +1 \] ,
\end{cases}
\end{eqnarray*}
 where $[t]$ denotes the integer part of $t$ except that $[t]=t-1$ when $t$ is an integer,  $\lambda$ is the time length of the trading period, $Z_t ^H= \int_{[t]}^t   dW_s$,  $Z_t ^L= \int_{\lambda+[t]}^t   dW_s$,  $\gamma= \gamma_H \gamma_L$, and $\theta = (\omega_{H1}, \omega_{H2}, \omega_L, \gamma_H, \gamma_L, \alpha_H,$ $\alpha_L, \beta_H, \beta_L, \nu_H, \nu_L )$ is the model parameters.
 For the jump part, $\Lambda_t$ is the Poisson process with constant intensity $\mu_{J}$, and the jump sizes  $J_t$'s are independent of  the continuous diffusion processes.
Furthermore,  the jump size $J_t$   is  equal to zero for the overnight period.
\end{definition}



The instantaneous volatility process of the OGI model is continuous with respect to time. 
That is, its trajectories are continuous. 
For the open-to-close period---for example, $ [t] \leq t \leq  [t] + \lambda$---the instantaneous volatility process reflects the market risk via the current integrated volatility and past squared overnight returns, whereas for the close-to-open period, $ [t]+\lambda \leq t \leq [t]+1$, the instantaneous volatility process utilizes the current log-return and past open-to-close integrated volatilities to express the market risk.
Specifically, the past risk factors are calculated through exponentially weighted averages with $\gamma$ order.
Furthermore, to account for the U-shape pattern of the intra-day volatility process \citep{andersen2018time},  the instantaneous volatility process has the quadratic terms with respect to time $t$. 
Thus, with appropriate choices of $\omega_{H1}$ and $\omega_{H2}$, the OGI model can explain the U-shape pattern.
At the market open time, the instantaneous volatility process has the following GARCH structure:
 \begin{eqnarray*}
 	\sigma_n^2 (\theta) &=& \omega_L +   \gamma_L  (\omega _{H1} -  \omega_{H2})  + \gamma  \sigma_{n-1}^2 (\theta) + \frac{\gamma_L \alpha_H}{  \lambda}\int_{n-1} ^{n-1 +\lambda} \sigma_t^2 (\theta) dt \cr
 	 &&+ \frac{\beta_L}{ 1-\lambda} (X_n - X_{n-1+\lambda} -\int_{n-1+\lambda}^{n} \mu_t dt  )^2, 
 \end{eqnarray*}
 where  $n$ is an integer,  
 and at the market close time, 
  \begin{eqnarray}\label{eq-SpotLF}
 	\sigma_{n+\lambda} ^2 (\theta) &=&   \omega_{H1} - \omega_{H2}  + \gamma_H \omega _L + \gamma  \sigma_{n-1+\lambda }^2 (\theta) + \frac{ \alpha_H}{   \lambda}\int_{n} ^{n+\lambda} \sigma_t^2 (\theta) dt  \cr
&&   + \frac{ \gamma_H  \beta_L }{ 1-\lambda} (X_n - X_{n-1+\lambda} -\int_{n-1+\lambda}^{n} \mu_t dt  )^2. 
 \end{eqnarray}
 Thus, the instantaneous volatility process is some  quadratic interpolation of the GARCH volatility with the open-to-close integrated volatility and squared close-to-open log-return as the innovation. 
To account for the random fluctuations of  the instantaneous volatilities, we introduce $Z_t ^H$ and $Z_t ^L$ with the scale parameters $\nu_H$ and $\nu_L$.
When considering only one of the open-to-close and close-to-open periods and ignoring the other period, the OGI model 
recovers the realized GARCH-It\^o process \citep{hansen2012realized, Song2018Volatlity} or unified GARCH-It\^o process \citep{kim2016unified}. 
Thus, unlike the proposed OGI model, these models only incorporate one innovation term of the integrated volatility and squared log-return in their conditional volatility.

Because our main interest lies in measuring the whole-day risk,   to estimate the model parameters,
we use nonparametric integrated volatility estimators 
 \citep{barndorff2008designing, jacod2009microstructure, zhang2006efficient} and squared log-returns as proxies for the parametric conditional expected integrated volatility.
Thus, it is important to investigate properties of the integrated volatility of the proposed OGI model. 
The following theorem shows the properties of integrated volatilities.
 
 \begin{theorem} 	\label{prop-integratedVol}
 For the OGI model,  we have the following properties. 

\begin{enumerate}
\item [(a)] The integrated volatilities have the following structure.
  For $0<\alpha_H<1$,  $0<\beta_L<1$, and $ n \in \mathbb{N}$, we have 
			\begin{eqnarray}
			&&\int_{n-1}^{n}  \sigma^2_t (\theta) dt = h_n (\theta) + D_n \quad \text{a.s.},  \label{model1}\\
			&&\int_{n-1} ^{ n-1 +\lambda}  \sigma_{t}^2(\theta)  dt = \lambda h_n^H (\theta)+D_n^H \quad a.s., \label{model2} \\
			&&\int_{ n-1 + \lambda} ^{n}  \sigma_{t}^2(\theta)  dt = (1-\lambda) h_n^L (\theta)+D_n^L \quad a.s.,  \label{model3}
		\end{eqnarray}
		where
			\begin{eqnarray*}	 
	&& h_n(\theta) = \omega^g + \gamma h_{n-1} (\theta) + \frac{\alpha  ^g}{ \lambda} \int_{n-2} ^{n-2+\lambda} \sigma_t^2 (\theta)dt  +\frac{\beta ^g}{ 1-\lambda } (X_{n-1} - X_{n-2+\lambda} - \int_{n-2+\lambda}^{n-1} \mu_t dt  ) ^2 ,  \label{eq3-con-GARCH}  \\
			&& h_n ^H (\theta) =  \omega_H^g + \gamma h_{n-1}^H (\theta) +  \frac{\alpha_H^g}{ \lambda} \int_{n-2}^{ n-2+\lambda} \sigma_t ^2 (\theta) dt + \frac{\beta_{H}^g}{ 1-\lambda } ( X_{n-1} - X_{n-2 +\lambda} - \int_{n-2+\lambda}^{n-1} \mu_t dt  )^2,  \cr
			&& h_n ^L (\theta) =  \omega_L^g + \gamma h_{n-1}^L (\theta) + \frac{\alpha_L^g}{ \lambda } \int_{n-2}^{ n-2+ \lambda} \sigma_t ^2 (\theta) dt +\frac{ \beta_{L}^g }{ 1-\lambda }( X_{n-1} - X_{ n-2+\lambda} - \int_{n-2+\lambda}^{n-1} \mu_t dt  )^2, 
			\end{eqnarray*}
			$D_n$, $D_n^H$, $D_n^L$ are martingale differences  and  $\omega^g, \gamma, \alpha^g, \beta^g,  \omega_H^g, \alpha_H^g, \beta_H^g,   \omega_L^g, \alpha_L^g, \beta_L^g$ are functions of $\theta$.
			 Their detailed forms are defined in   Theorem \ref{prop-integratedVol-apendix} in the supplement document.

\item [(b)] We have
  \begin{eqnarray*}
 	&&E \[ \(D_n^H \) ^2 \middle | \FF_{n-1} \] = \varphi_{n}^H (\theta)    =   \lambda^{2} \nu_H^g      \text{ a.s.}, \cr
 	&&E \[ \(D_n^{LL} \) ^2 \middle | \FF_{n-1} \] = \varphi_{n}^L (\theta )  \cr
 	&&   =   F_{\beta_L, 1} s_{n-1}^4 (\theta) +  F_{\beta_L,2}  \omega_L s_{n-1}^2 (\theta) +  F_{\beta_L,3} \omega_L^2     +   (1-\lambda)^2 \nu_L^g     \text{ a.s.} ,
 \end{eqnarray*}
  where  $ D_n^{LL} = D_n^L +2 \int_{\lambda +n-1}^ n (X_t - X_{\lambda + n-1} ) \sigma_t (\theta_0) dB_t $, $\nu_H^g$ and $\nu_L^g$ are defined in \eqref{eq-nu} and \eqref{eq-nu1}, respectively,  $s_{n-1}^2 (\theta) $ is defined in \eqref{eq-s}, and $ F_{\beta_L, i}$'s are   functions of $\beta_L$ defined in \eqref{F-def} in the supplement document.
\end{enumerate} 
\end{theorem}

 Theorem \ref{prop-integratedVol} (a) shows that the integrated volatility can be decomposed into the GARCH volatility and martingale difference.
 This structure implies that the daily conditional expected volatility is a function of the past open-to-close integrated volatilities  and squared close-to-open log-returns.
That is, under the OGI process, the market dynamics can be explained by the open-to-close integrated volatility and squared close-to-open log-returns, which represent volatilities for the open-to-close and close-to-open periods, respectively. 
Thanks to these two different volatility sources, we expect the proposed OGI model to capture the market dynamics well.   
In the empirical study, we find  that the integrated volatilities and squared log-returns help to account for the market dynamics (see Section \ref{SEC-empirical}).

 \begin{remark}
 Based on the result of Theorem \ref{prop-integratedVol}, we can predict the one period ahead volatility.
 In practice, we often need to predict the multi-period ahead volatility.  
To do this, we use the following relationship:
\begin{eqnarray*}
	 E  \left [ \int_{n-1}^{n}  \sigma^2_t (\theta) dt  \middle | \mathcal{F}_{n-2}  \right ]  &=&  	E  \left [  E  \left [  \int_{n-1}^{n}  \sigma^2_t (\theta) dt   \middle | \mathcal{F}_{n-1}  \right ]   \middle | \mathcal{F}_{n-2}  \right ]  \cr
	& =& E  \left [  h_{n} (\theta)   \middle | \mathcal{F}_{n-2}  \right ]  \cr
	& =&  \omega^g + \gamma h_{n-1} (\theta) +  \alpha  ^g  h_{n-1}^H (\theta)  + \beta ^g  h_{n-1}^L (\theta)  \text{ a.s.}
\end{eqnarray*}
Then, recursively, we can obtain the multi-period prediction. 
 \end{remark}

As we discussed above, we  estimate the model parameters via the relationship between the conditional GARCH volatilities, $h_n^H(\theta)$, $h_n ^L (\theta)$, and $h_n (\theta)$, and the corresponding integrated volatility  or squared log-return.
Thus, to study the low-frequency volatility dynamics, we only need  Theorem \ref{prop-integratedVol} (a).
That is, under the model assumptions \eqref{model1}--\eqref{model3}, we develop the rest of the paper.
In comparison with direct volatility modeling based on realized volatility such as HAR, HEAVY, and  realized GARCH models 
 \citep{andersen2003modeling, corsi2009simple, hansen2012realized, shephard2010realising}, the unified GARCH-It\^o model and 
 OGI model may be more difficult or even less practical for drawing statistical inferences from combined low- and high-frequency data. 
However, like the unified GARCH-It\^o model case, the OGI approach indicates 
the existence of the diffusion process, which satisfies the conditions \eqref{model1}--\eqref{model3} and fills the gap between the low-frequency discrete time series volatility modeling and the high-frequency continuous time  diffusion process.
 Because the purpose of this paper is to develop diffusion processes that can account for the low-frequency market dynamics, the parameter of interest is the GARCH parameter $\theta^g= (\omega_{H}^g, \omega_{L}^g, \gamma,  \alpha_H ^g, \alpha_L^g, \beta_{H}^g  , \beta_{L}^g )$.
We notice that, under the model assumption, we need the common $\gamma$ condition for the open-to-close and close-to-open conditional volatilities to have  the GARCH conditional volatility form for  $h_n^H(\theta)$, $h_n ^L (\theta)$, and $h_n (\theta)$. 
When it comes to estimating GARCH parameters, we assume that the open-to-close and close-to-open volatilities have different dynamic structures, so we make inferences for $h_n^H(\theta)$ and $h_n ^L (\theta)$ separately under the common $\gamma$ condition. 
Details can be found in Section  \ref{SEC-WLSE}.

 \section{Estimation procedure} \label{SEC-WLSE}
  
 \subsection{A model setup}
 We assume that the underlying diffusion process follows the OGI process defined in Definition \ref{Def-1}. 
 The high-frequency observations during the $d$th open-to-close period are observed at $t_{d,i}, i=1,\ldots, m_d$, where $d-1=t_{d,0} < t_{d,1} < \cdots < t_{d, m_d} = \lambda+ d-1$.  
 Let $m$ be the average number of the high-frequency observations, that is, $m= \frac{1}{n} \sum_{d=1}^n m_d$.  
Due to market inefficiencies, such as the bid-ask spread, asymmetric information, and so on,  the high-frequency data are masked by the microstructure noise.
 To account for this, we assume that the observed log-prices during the open-to-close period have the following additive noise structure:
 \begin{equation*}
 	Y_{t_{d,i}}= X_{t_{d,i}} + \epsilon_{t_{d,i}}, \quad \text{for } d=1,\ldots, n, i=1,\ldots, m_d-1, 
 \end{equation*}
 where $X_t$ is the true log-price, $\epsilon_{t_{d,i}}$ is microstructure noise with mean zero and variance $\eta_d$, and the log-price and microstructure noise are independent.
The effect of $\mu_t$ is negligible regarding high-frequency realized volatility estimators, and  the magnitude of daily returns is relatively small.
Thus, for simplicity, we assume $\mu_t=0$ in Definition \ref{Def-1}. 
We note that the theoretical results in Theorem \ref{Thm-1} can be established in the similar way with non-zero $\mu_t$ under some piecewise constant condition for $\mu_t$. 
In contrast,  during the close-to-open period, we only observe the low-frequency observations, open and close prices. 
In the low-frequency time series modeling, we often assume that the true low-frequency prices are observed.
In practice, the microstructure noise may exist in the low-frequency observations, but its impact on  the low-frequency modeling is relatively small. 
Thus, we also assume that the true low-frequency observations, the open and close prices $X_d$ and $X_{\lambda+d}$, are observed at the open and close times, $t_{d+1,0}$ and $t_{d+1,m_{d+1}}$.

\begin{remark}
For the microstructure noise, we may need a stationary condition to estimate the integrated volatility with the optimal convergence rate $m^{-1/4}$ \citep{  barndorff2008designing,  fan2018robust, jacod2009microstructure, kim2016asymptotic, 
zhang2006efficient}.  For example, we may impose a ARMA-type structure on the microstructure noise and assume some dependence 
between the price processes and the microstructure noise. 
However, in this paper, we directly adopt a well-performing nonparametric realized volatility estimator, which can be obtained under certain structures of the microstructure without affecting the volatility modeling.   
Thus, we can put such structures on the microstructure noise, as long as we can secure the well-performing realized volatility estimator. 
\end{remark}

%

 \subsection{GARCH parameters estimation}
 We first fix some notations. 
 For any given vector $b= (b_i) _{i=1,\ldots, k}$, we define $\|b \|_{\max}=\max_i |b_i|$.
 Let $C$'s be positive generic constants whose values are
independent of $\theta$, $n$, and $m$  and may change from occurrence to occurrence. 
 In this section, we develop an estimation procedure for  the GARCH parameters, $\theta^g= (\omega_{H}^g, \omega_{L}^g, \gamma,  \alpha_H ^g, \alpha_L^g, \beta_{H}^g  , \beta_{L}^g )$,  which are minimum required parameters to evaluate the GARCH volatilities defined in Theorem \ref{prop-integratedVol}, where elements of $\theta^g$ are defined in Theorem \ref{prop-integratedVol-apendix} in the supplement document.
We denote the  true GARCH parameter by $\theta_0^g  = (\omega_{H,0}^g, \omega_{L,0}^g, \gamma_{0},  \alpha_{H,0} ^g, \alpha_{L,0}^g, \beta_{H,0}^g  , \beta_{L,0}^g )$.

Theorem \ref{prop-integratedVol} indicates that integrated volatilities can be decomposed into the GARCH volatility terms $h_n^H (\theta_0^g)$ and $h_n^L (\theta_0^g)$, and the martingale difference terms $D_n^H$ and $D_n^L$.
This fact inspires us to use the integrated volatilities as proxies of the GARCH volatilities.
Then, as the sample period goes to infinity, the martingale convergence theorem may provide consistency of the estimators. 
However, the integrated volatilities are not observable, so we first need to estimate them.
 For the open-to-close period, we use the high-frequency observations to estimate the open-to-close integrated volatility nonparametrically \citep{
 ait2012testing,  barndorff2008designing, corsi2010threshold, fan2007multi,  jacod2009microstructure,  xiu2010quasi, zhang2006efficient, zhang2016jump},  and we call these nonparametric estimators ``realized volatility."
  Under mild conditions, we can show that realized volatility converges to integrated volatility with the optimal convergence rate $m^{-1/4}$ \citep{barndorff2008designing, jacod2009microstructure, kim2016asymptotic, tao2013fast, xiu2010quasi, zhang2006efficient}. 
 In the numerical study, we employ the jump robust pre-averaging realized (PRV) estimator \citep{ait2016increased, jacod2009microstructure}.
 However, for the close-to-open period, high-frequency data are not available, so we use the squared close-to-open return as the proxy. 
 Note that It\^o's lemma indicates 
 $$
 (X_n -X_{\lambda +n-1} ) ^2= \int_{\lambda + n-1} ^n  \sigma_t ^2 (\theta_0)dt + 2 \int_{\lambda +n-1}^ n (X_t - X_{\lambda + n-1} ) \sigma_t (\theta_0) dB_t \text{ a.s.}
 $$
This implies that the squared close-to-open return can also be decomposed into the GARCH volatility and martingale difference. 
That is, we have the following relationships:
\begin{eqnarray*}
&& \int_{n-1} ^{\lambda+n-1}  \sigma_t ^2 (\theta_0)dt  = \lambda h_{n}^H (\theta_0^g) + D_{n}^H \text{ a.s.}, \cr
&& (X_n -X_{\lambda +n-1} ) ^2 = (1-\lambda) h_{n}^L (\theta_0^g) + D_{n}^{LL} \text{ a.s.},
\end{eqnarray*}
where $ D_n^{LL} =D_n^L +2 \int_{\lambda +n-1}^ n (X_t - X_{\lambda + n-1} ) \sigma_t (\theta_0) dB_t $.
We use the above relationships to estimate the GARCH parameter  $\theta_0^g$.

The variances of  the martingale differences $D_{n}^H$ and $D_{n}^{LL}$ indicate the accuracy of the GARCH volatility information coming from the proxies $\int_{n-1} ^{\lambda+n-1}  \sigma_t ^2 (\theta_0)dt$ and $(X_n -X_{\lambda +n-1} ) ^2$, so each proxy with the smaller variance is closer to the corresponding GARCH volatility. 
Thus, as we incorporate the variance information into an estimation procedure, we expect to improve its performance.
For example,  we can standardize the proxies as follows:
\begin{equation*}
  \frac{ \(\int_{n-1} ^{\lambda+n-1}  \sigma_t ^2 (\theta_0)dt   - \lambda h_{n}^H (\theta_0^g) \) ^2}{ E \[ \(D_{n}^H \)^2\]} \quad   \text{ and } \quad   \frac{\( (X_n -X_{\lambda +n-1}  ) ^2-  (1-\lambda) h_{n}^L (\theta_0^g) \)^2 } {E \[ \(D_{n}^{LL} \)^2 \]  }.
\end{equation*}
The unit expectations help to assign a larger weight to a more accurate proxy. 
In the empirical study, we find that the variance of the integrated volatilities is smaller than that of the squared  close-to-open returns.
That is, the open-to-close proxy is more accurate, so we make more use of the information from the open-to-close period by assigning to it 
a larger weight. 
To compare the proxies and GARCH volatilities, we employ 
the weighted least squares estimation as follows:
  \begin{eqnarray*}
 	L_{n} (\theta^g)  &=&  -\frac{1}{n} \sum_{i=1}^n  \Bigg [     \frac{ (IV_i-  \lambda h_i  ^H(\theta^g ))^2 }{ \hat{\phi}_H }       + \frac{ \( ( X_{i} - X_{\lambda+ i-1}   )^2-(1-\lambda) h_i ^L  (\theta^g )  \)^2  }{ \hat{\phi}_L   } \Bigg ],
 \end{eqnarray*}
 where the GARCH volatility terms $h_i  ^H(\theta^g )$ and $h_i  ^L(\theta^g )$ are defined in Theorem \ref{prop-integratedVol}, $IV_i=    \int_{i-1}^{\lambda+i-1} \sigma_t ^2(\theta_0)dt$, and $\hat{\phi}_H$  and $\hat{\phi}_L$ are consistent estimators of variances of martingale differences $D_n^H$ and $D_{n}^{LL}$, respectively.
To evaluate the above quasi-likelihood function, we first need to estimate the integrated volatility $IV_i$. 
It can be estimated by  the realized volatility estimator, which is denoted by $RV_i$. 
Then we estimate the GARCH volatilities as follows: 
 \begin{eqnarray}
 			&& \hat{h}_n ^H (\theta^g) =  \omega_H^g + \gamma  \hat{h}_{n-1}^H (\theta^g) + \frac{\alpha_H^g }{\lambda }RV_{n-1} + \frac{\beta_{H}^g}{  1-\lambda } ( X_{n-1} - X_{\lambda+ n-2} )^2,  \label{eq-quantity1} \\
 			&&\hat{h}_n ^L (\theta^g) =  \omega_L^g + \gamma \hat{h}_{n-1}^L (\theta^g) + \frac{\alpha_L^g}{ \lambda } RV_{n-1}+ \frac{\beta_{L}^g}{  1-\lambda }( X_{n-1} - X_{\lambda+ n-2} )^2 \label{eq-quantity2}.
  \end{eqnarray}
  We note that the conditional expected volatilities for the open-to-close and close-to-open periods have the common $\gamma$ as in \eqref{eq-quantity1} and \eqref{eq-quantity2}, which makes it possible to have the GARCH form for the whole-day conditional expected volatility.  
  To evaluate the GARCH volatilities, we use $RV_1$ and the sample variance of the close-to-open log-returns as the initial values $h_0^H(\theta^g)$ and $h_0^L(\theta^g)$, respectively. 
The effect of the initial value has the negligible order $n^{-1}$ (see Lemma 1 in \citet{kim2016unified}),  so its choice does not significantly affect the parameter estimation.  
  With these estimators, we define the quasi-likelihood function as follows:
  \begin{eqnarray}\label{eq-QL}
 	\hat{L}_{n,m} (\theta^g) & =&  -\frac{1}{n} \sum_{i=1}^n \Bigg [      \frac{ (RV_i-  \lambda \hat{h}_i  ^H(  \theta^g  ))^2 }{\hat{\phi}_H   } +    \frac{ \( ( X_{i} - X_{\lambda+ i-1} )^2-(1-\lambda) \hat{h}_i ^L  (\theta^g)  \)^2  }{ \hat{\phi}_L} \Bigg ],
 	\end{eqnarray}
  and we obtain the estimator of the GARCH parameters $\theta_0^g$ by maximizing the quasi-likelihood function. That is,
 \begin{equation*}
	\hat{\theta}^g = \arg \max _{\theta^g  \in \Theta^g } \hat{L}_{n,m} (\theta^g ),
 \end{equation*}
 where $\Theta^g$ is the parameter space of $\theta^g$.
 We call the estimator the weighted least squares estimator (WLSE).
 To obtain the variances of martingale differences, $ \hat{\phi}_H  $ and $ \hat{\phi}_L $, we employ the QMLE method as follows.
 We define the quasi-likelihood functions for the open-to-close and close-to-open, respectively, in the following manner: 
   \begin{eqnarray} 
 	&& \hat{L}_{n,m}^H (\theta_H^g) =  -\frac{1}{n} \sum_{i=1}^n  \[  \log (\lambda \hat{h}_i  ^H(  \theta_H^g  )) + \frac{  RV_i  }{\lambda \hat{h}_i  ^H(  \theta_H^g  )   }  \] , \label{sep-model1} \\
 	&& \hat{L}_{n,m}^L (\theta_L^g)  =  -\frac{1}{n} \sum_{i=1}^n   \[ \log ((1-\lambda) \hat{h}_i ^L  (\theta_L^g) ) +  \frac{   ( X_{i} - X_{\lambda+ i-1} )^2 }{(1-\lambda) \hat{h}_i ^L  (\theta_L^g)  }   \], \label{sep-model2} 
 	\end{eqnarray}
 	where $\theta_H^g=   (\omega_{H}^g,  \gamma,  \alpha_H ^g, \beta_{H}^g )$ and $\theta_L^g=   ( \omega_{L}^g, \gamma,   \alpha_L^g, \beta_{L}^g )$.
 	Then we find their maximizers, which are denoted by $\hat{\theta}_H^g$ and $\hat{\theta}_L^g$. 
Using the residuals, we estimate the variances of  martingale difference in the following way:
 \begin{eqnarray*}
 	&&\hat{\phi}_H= \frac{1}{n} \sum_{i=1}^n  (RV_i -\lambda \hat{h}_i  ^H(  \hat{\theta}_H^g  )) ^2, \cr
 	&&\hat{\phi}_L= \frac{1}{n} \sum_{i=1}^n(   ( X_{i} - X_{\lambda+ i-1} )^2- (1-\lambda) \hat{h}_i ^L  (\hat{\theta}_L^g) )^2.
 \end{eqnarray*} 
 Similar to the proofs of Theorems 3 and 5 in \citet{kim2016unified}, we can establish their consistency. 
 
 \begin{remark}
 There are other possible choices of the  variance of the martingale differences. 
 For example, we can use the conditional variances in Theorem \ref{prop-integratedVol} (b) to evaluate the quasi-likelihood function \eqref{eq-QL}. 
 However, the conditional variance heavily depends on the underline OGI process, which may cause some bias when the underline model is misspecified. 
 Thus, to make robust inferences, we use the unconditional variance instead of the conditional variance. 
 Furthermore, the proposed procedure has a more simple structure, which may help to reduce estimation errors. 
  We note that the proposed two-step weighted least square estimation procedure works well as long as the first-step variance estimators are consistent.
 Thus, we can easily incorporate the other variance estimator. 
 According to our empirical analysis, the unconditional variance estimator provides more stable results than the conditional variance estimator. 
 Thus, we use the unconditional variance and only report its related results. 
 If we can estimate conditional  variance in a robust way, it may show better performance. 
 However, obtaining the robustness is not straightforward because we need to impose structure on the process to evaluate the conditional variance.  
 We leave this for a future study.  
   \end{remark}

 To establish asymptotic properties for the proposed WLSE, we need the following technical assumptions.

 \begin{assumption} \label{Assumption1}
~
\begin{itemize}

\item [(1)] $ \theta_0^g  \in \Theta^g  = \{  (\omega_H^g, \omega_L^g , \gamma, \alpha_{H}^g, \alpha_{L}^g, \beta_H^g, \beta_L^g  ); \omega_l  < \omega _H^g, \omega_L ^g < \omega_u  , \gamma_l  < \gamma < \gamma_u  <1, \alpha_l < \alpha_H^g, \alpha_L^g < \alpha_u,  \beta_l  < \beta_H^g, \beta_L^g < \beta_u <1, \| \bA \|_2 <1  \}$, 
where $\omega_l$, $\omega_u$, $\gamma_l$, $\gamma_u$, $\alpha_l$, $\alpha_u$, $\beta_l$, $\beta_u$  are some known positive constants, $\| \cdot \|_2$ is the matrix spectral norm, and 
$
\bA =  \begin{pmatrix}
  \gamma + \frac{\alpha_{H}^g }{\lambda} & \frac{\beta_{H}^g}{1-\lambda}
\\ 
\frac{\alpha_{L} ^g}{ \lambda} & \gamma + \frac{\beta_{L}^g}{1-\lambda}
\end{pmatrix}.
$ 

\item[(2)] We have for some positive constant $C$,
\begin{equation*}
  \sup_d E \[  (X_{\lambda+d-1} - X_{d-1}) ^{4}  \] \leq C,  \quad \sup_d E \[  (X_{d} - X_{\lambda+d-1}) ^{4}  \] \leq C,  \quad  \sup_d E \[   \(D_{d}^{LL} \) ^{4} \] \leq C.
\end{equation*}

 \item[(3)]   We have $C_1 m \leq m_d \leq C_2 m$ for all $d$, and  $\max_d \sup _{1 \leq j \leq m_d} |t_{d,j} -t_{d,j-1}| =O(m^{-1})$.

\item [(4)]  For any $d \in \mathbb{N}$, $E( |RV_d  - IV_d |^2) \leq C m^{-1/2}$ .

\item[(5)]  $( D_{d} ^H,  D_{d}^{LL} )$ is a stationary process.

\item[(6)]  $|\hat{\phi}_H -\phi_{H0} | = o_p(1) $ and $|\hat{\phi}_L -\phi_{L0}| = o_p(1) $, where $\phi_{H0} =E \[ \( D_n ^{H}\)^2\]$ and $\phi_{L0} =E\[ \( D_n ^{LL}\) ^2 \]$. 
\end{itemize}
 \end{assumption}
 
 \begin{remark}
 Assumption \ref{Assumption1}(2) is about the finite 4th moment condition,  which is the minimum requirement when handling the second moment target parameter. 
 Under some finite 4th moment conditions,  Assumption \ref{Assumption1}(4) is satisfied \citep{kim2018large, tao2013fast}.
However, when there is a jump part in the diffusion process, this condition  may be violated. 
In this case, we need to employ some jump robust realized volatility \citep{ait2016increased, zhang2016jump} and derive some uniform  convergence with respect to time $d$. 
 Finally, Assumption \ref{Assumption1}(5) is required to derive an asymptotic normal distribution of the proposed WLSE. 
 \end{remark}

  The following theorem investigates the asymptotic behaviors of the proposed WLSE  $\hat{\theta}$.

 \begin{theorem}\label{Thm-1}
 Under Assumption \ref{Assumption1}, we have
 \begin{equation}\label{eq-result1-Thm2}
 	\|\hat{\theta}^g  - \theta _0^g \|_{\max} = O_p ( m^{-1/4} + n^{-1/2}). 
 \end{equation}
Furthermore, we suppose that $n m^{-1/2} \rightarrow 0$ as $m, n \rightarrow \infty$.
Then we have 
 \begin{equation}\label{eq-result2-Thm2}
 	\sqrt{n} ( \hat{\theta}^g  - \theta _0^g) \overset{d}{\to}   N (0,  A^{-1}B A^{-1}) ,
 \end{equation}
 where 
 \begin{eqnarray*}
 	&&A = E \Bigg [  \frac{  \lambda^2}{ \phi_{H0} } \frac{\partial h_1^H (\theta^g ) }{ \partial \theta ^g}\frac{\partial h_1^H (\theta^g ) }{ \partial \theta^{g\top} } \Big | _{\theta^g  = \theta _0^g}   + \frac{  (1- \lambda)^2}{ \phi_{L0}  } \frac{\partial h_1^L (\theta^g) }{ \partial \theta^g}\frac{\partial h_1^L (\theta^g) }{ \partial \theta^{g \top} } \Big| _{\theta^g  = \theta _0^g}  
 \Bigg ],  \cr
 && B =   E \Bigg [     \frac{  \lambda^2 \varphi_{1}^H (\theta_0)   }{ \phi_{H0} ^2 } \frac{\partial h_1^H (\theta^g ) }{ \partial \theta ^g}\frac{\partial h_1^H (\theta^g ) }{ \partial \theta^{g\top} } \Big | _{\theta^g  = \theta _0^g}   + \frac{  (1- \lambda)^2 \varphi_{1}^L (\theta_0) }{ \phi_{L0}^2  } \frac{\partial h_1^L (\theta^g) }{ \partial \theta^g}\frac{\partial h_1^L (\theta^g) }{ \partial \theta^{g \top} } \Big| _{\theta^g  = \theta _0^g}  
 \Bigg ], 
 \end{eqnarray*}
 $\varphi_{i}^H (\theta_0)$ and  $\varphi_{i}^L (\theta_0)$ are defined in Theorem  \ref{prop-integratedVol}(b).
   \end{theorem}

 \begin{remark}
 Theorem \ref{Thm-1} shows that the WLSE $\hat{ \theta}^g$ has the convergence rate $m^{-1/4} + n^{-1/2}$. 
 The first term, $m^{-1/4}$,  comes from estimating the integrated volatility, which is known as the optimal convergence rate in the case of 
 high-frequency data with the presence of the microstructure noise. 
 The second term, $n^{-1/2}$, is the usual convergence rate in the low-frequency data case. 
 Under the stationary assumption, we also derive the asymptotic normality. 
 \end{remark}
 
\begin{remark}
To derive the asymptotic normality, we need the condition $n m^{-1/2} \rightarrow 0$, which is too restrictive for the long sample period. 
If this condition is violated, the asymptotic normality may depend on $m^{1/4} (RV_d  - IV_d)$, which is the quantity related to high-frequency estimation. 
If this term is some martingale difference, we may be able to relax the condition such as $n m^{-1} \rightarrow 0$. In this case, usually, $m$ is huge, so it is not restrictive. 
\end{remark}

 One of our objectives in this paper is to predict future volatility. 
 The best predictor given the current available information $\FF_n$ is the conditional expected volatility---that is, the GARCH volatility $h_{n+1}(\theta_0^g)$. 
With the model parameter estimator, we estimate the GARCH volatility as follows:
 \begin{equation*}
 	\hat{h}_{n+1} (\hat{\theta^g}) = \hat{\omega}^g +  \hat{\gamma} \hat{h}_{n} (\hat{\theta}^g) +\hat{\alpha}^g  \lambda^{-1} RV_n + \hat{\beta}^g (1-\lambda)^{-1} (X_n - X_{\lambda+n-1}) ^2,
 \end{equation*} 
 where the GARCH parameters $ \hat{\omega}^g$, $\hat{\alpha}^g$,  and $\hat{\beta}^g$ are estimated using the plug-in method with the WLSE $\hat{\theta}^g$. 
The following corollary provides the consistency of the GARCH volatility estimator.

\begin{corollary}
Under the assumptions of Theorem \ref{Thm-1} (except for $n m^{-1/2} \rightarrow 0$), we have
\begin{equation*}
	|\hat{h}_{n+1} (\hat{\theta}^g)  - h_{n+1} (\theta_0^g) | = O_p (n^{-1/2} + m^{-1/4}).
\end{equation*}
\end{corollary}

  \subsection{Hypothesis tests} \label{SEC-Normality}

 In financial practices, we are interested in the GARCH parameters $(\omega^g, \gamma, \alpha^g, \beta^g)$ and often make statistical inferences about them, such as hypothesis tests.
 In this section, we discuss how to conduct hypothesis tests for the GARCH parameters.

  We first derive the asymptotic distribution of the GARCH parameter estimators.
  Theorem \ref{Thm-1} implies that 
    \begin{equation*} 
 	\sqrt{n}  ( \hat{\theta} ^g - \theta _0^g) \overset{d}{\to}   N (0, A^{-1} B A^{-1} )  ,
 \end{equation*}
 where 
 \begin{eqnarray*}
 	&&A = E \Bigg [  \frac{  \lambda^2}{ \phi_{H0} } \frac{\partial h_1^H (\theta^g ) }{ \partial \theta ^g}\frac{\partial h_1^H (\theta^g ) }{ \partial \theta^{g\top} } \Big | _{\theta^g  = \theta _0^g}   + \frac{  (1- \lambda)^2}{ \phi_{L0}  } \frac{\partial h_1^L (\theta^g) }{ \partial \theta^g}\frac{\partial h_1^L (\theta^g) }{ \partial \theta^{g \top} } \Big| _{\theta^g  = \theta _0^g}  
 \Bigg ],  \cr
 && B =   E \Bigg [     \frac{  \lambda^2 \varphi_{1}^H (\theta_0)   }{ \phi_{H0} ^2 } \frac{\partial h_1^H (\theta^g ) }{ \partial \theta ^g}\frac{\partial h_1^H (\theta^g ) }{ \partial \theta^{g\top} } \Big | _{\theta^g  = \theta _0^g}   + \frac{  (1- \lambda)^2 \varphi_{1}^L (\theta_0) }{ \phi_{L0}^2  } \frac{\partial h_1^L (\theta^g) }{ \partial \theta^g}\frac{\partial h_1^L (\theta^g) }{ \partial \theta^{g \top} } \Big| _{\theta^g  = \theta _0^g}  
 \Bigg ]. 
 \end{eqnarray*}
 The GARCH parameters are functions of $\theta$.
 For example, 
 $\omega^g =  \lambda  \omega_H^g + (1-\lambda) \omega_L ^g$, $\alpha ^g = \lambda   \alpha_H^g + (1-\lambda) \alpha_L^g $,  $\beta  ^g = \lambda   \beta_{H}^g  +  (1-\lambda)\beta_{L}^g$, where $\alpha_H^g, \alpha_L^g, \beta_H^g, \beta_L^g$ are defined in Theorem \ref{prop-integratedVol-apendix}.
 Thus, using the delta method and Slutsky's theorem, we can show that when $ 	\frac{\partial f(\theta^g) }{\partial \theta^g}   | _{\theta^g= \theta_0^g} \neq 0$, 
  \begin{equation}\label{Z-stat}
T_{f,n}= \frac{\sqrt{n}  (f(\hat{\theta}^g) - f(\theta_0^g))}{  \sqrt{  \triangledown f (\hat{\theta}^g)  ^{\top} (\hat{A}^{-1} \hat{B} \hat{A}^{-1} ) ^{-1} \triangledown f (\hat{\theta}^g)    }} \overset{d}{\to}   N (0, 1), 
  \end{equation}
  where $  \triangledown f (\hat{\theta}^g) =	\frac{\partial f(\theta^g) }{\partial \theta^g}   | _{\theta^g= \hat{\theta}^g}$ and $\hat{A}$ and $\hat{B}$ are  consistent estimators of $A$ and $B$, respectively.
   To evaluate the asymptotic variances of the GARCH parameter estimators, we first need to estimate $A$ and $B$. 
We use the following estimators,
$$
  \hat{A} (\theta^g ) = - \frac{\partial ^2 \hat{L} _{n,m} (\theta ^g) }{ \partial \theta^g \partial \theta^{g\top} } \quad \text{and} \quad \hat{B}(\theta^g) = \frac{1}{n} \sum_{i=1}^n \frac{\partial \hat{l}_{i} (\theta^g)}{\partial \theta^g}\frac{\partial 
\hat{l}_{i} (\theta^g)}{\partial \theta ^{g\top}},
$$
and
$$
\hat{l}_{i} (\theta^g) = \frac{ (RV_i-  \lambda \hat{h}_i  ^H(\theta^g ))^2 }{ \hat{\phi}_{H} }       +     \frac{ \( ( X_{i} - X_{\lambda+ i-1}   )^2-(1-\lambda) \hat{h}_i ^L  (\theta^g)  \)^2  }{\hat{\phi}_L   },
$$
and  $\hat{h}_i^H(\theta^g)$, $\hat{h}_i^L(\theta^g)$  are defined in \eqref{eq-quantity1} and \eqref{eq-quantity2}, respectively.
  Under some stationary condition, we can establish its consistency. 
Then, using the proposed Z-statistics $T_{f,n}$ in \eqref{Z-stat}, we can conduct the hypothesis tests based on the standard normal distribution.

 \section{A simulation study}\label{SEC-simulation}
 
 We conducted simulations to check the finite sample performance of the proposed estimation methods.
 We generated the log-prices for $n$ days with frequency $1/m^{all}$ for each day and let $t_{d,j}= d-1+ j/m^{all}$, $d=1,\ldots, n, j= 0, \ldots, m^{all}$.  We chose the closed time $\lambda$ as $6.5/24$, which corresponds to 6.5 trading hours. 
The true log-price follows the OGI model in Definition \ref{Def-1}. 
The parameter setup is presented in the supplement document. 
To generate the jump, we simply set the jump size as $|J_t| =0.05$ and the signs of the jumps were randomly generated.
$\Lambda_t$ was generated using a Poisson distribution with mean 10 during the open-to-close period. 
 For the open-to-close period, we generated the noisy observations.
The detail setup can be found in the supplement document.
 To generate the true process, we chose $m^{all} = 43,200$, which equals the number of every 2 seconds in a one-day period.
 We varied $n$ from 100 to 500 and $m$ from  390 to 11,700, which correspond to the numbers of 1 minute  and every 2 seconds in the open-to-close period, respectively. 
  We repeated the whole procedure 500 times.

 \begin{table}[!ht] \small
\begin{center}
\caption{Mean absolute errors (MAE) for the WLSE estimates with $n=100, 200, 500$ and $m=  390, 1170, 2340$ and the true vol.}%
\begin{tabular*}{0.8 \textwidth}{@{\extracolsep{\fill}}cc cccccccc}
 \hline
 &&  \multicolumn{7}{c}{MAE$\times 10^2$} \\ \cline{3-9} 
$n$	&	$m$	&	$\omega_H^g$	&	$\omega_L^g$	&	$\gamma$	&	$\alpha_H^g$	&	$\alpha_L^g$	&	$\beta_H^g$	&	$\beta_L^g$	\\ \hline
100	&	390	&	0.0383 	&	0.0497 	&	0.2329 	&	0.1627 	&	0.2145 	&	0.0455 	&	0.1014 	\\
	&	1170	&	0.0383 	&	0.0498 	&	0.2325 	&	0.1626 	&	0.2141 	&	0.0453 	&	0.1013 	\\
	&	2340	&	0.0291 	&	0.0484 	&	0.1696 	&	0.1064 	&	0.2307 	&	0.0299 	&	0.1024 	\\
	&	True vol	&	0.0231 	&	0.0447 	&	0.1483 	&	0.0817 	&	0.2143 	&	0.0233 	&	0.1023 	\\
	&		&		&		&		&		&		&		&		\\
200	&	390	&	0.0247 	&	0.0338 	&	0.1619 	&	0.1444 	&	0.1593 	&	0.0380 	&	0.0685 	\\
	&	1170	&	0.0247 	&	0.0338 	&	0.1615 	&	0.1445 	&	0.1593 	&	0.0378 	&	0.0685 	\\
	&	2340	&	0.0191 	&	0.0343 	&	0.1091 	&	0.0781 	&	0.1645 	&	0.0229 	&	0.0730 	\\
	&	True vol	&	0.0136 	&	0.0307 	&	0.0880 	&	0.0513 	&	0.1583 	&	0.0164 	&	0.0749 	\\
	&		&		&		&		&		&		&		&		\\
500	&	390	&	0.0159 	&	0.0230 	&	0.1191 	&	0.1329 	&	0.1171 	&	0.0329 	&	0.0446 	\\
	&	1170	&	0.0159 	&	0.0230 	&	0.1191 	&	0.1330 	&	0.1173 	&	0.0328 	&	0.0446 	\\
	&	2340	&	0.0134 	&	0.0235 	&	0.0726 	&	0.0586 	&	0.1079 	&	0.0190 	&	0.0490 	\\
	&	True vol	&	0.0082 	&	0.0212 	&	0.0545 	&	0.0330 	&	0.1053 	&	0.0095 	&	0.0519 	\\ \hline

	\end{tabular*}
\label{Table-1}
\end{center}
\end{table}

 Table \ref{Table-1} reports the mean absolute errors (MAE),  $|\hat{\theta} - \theta_0|$ of the WLSE estimates   with $n=100, 200, 500$ and $m=  390, 1170, 11700$ and the true vol. 
 From  Table \ref{Table-1},  we find that the mean absolute errors usually decrease as the number of high-frequency observations or low-frequency observations increases. 
 The close-to-open period parameters, $\omega_L^g$, $\alpha_L^g$, $\beta_L^g$, are mainly dependent on the number of  low-frequency observations, whereas the open-to-close period parameters, $\omega_H^g$, $\gamma$, $\alpha_H^g$, $\beta_H^g$, are dependent on the number of both the high-frequency and low-frequency observations. 
  This is because  the close-to-open period parameters are estimated based on only the low-frequency data, whereas the  open-to-close period parameters, $\omega_H^g$, $\gamma$, $\alpha_H^g$, $\beta_H^g$, are estimated based on both the high-frequency and low-frequency data. 
    This result supports the theoretical findings in Section \ref{SEC-WLSE}.

  \begin{figure}[!ht]
  \centering
    \includegraphics[width=1\textwidth]{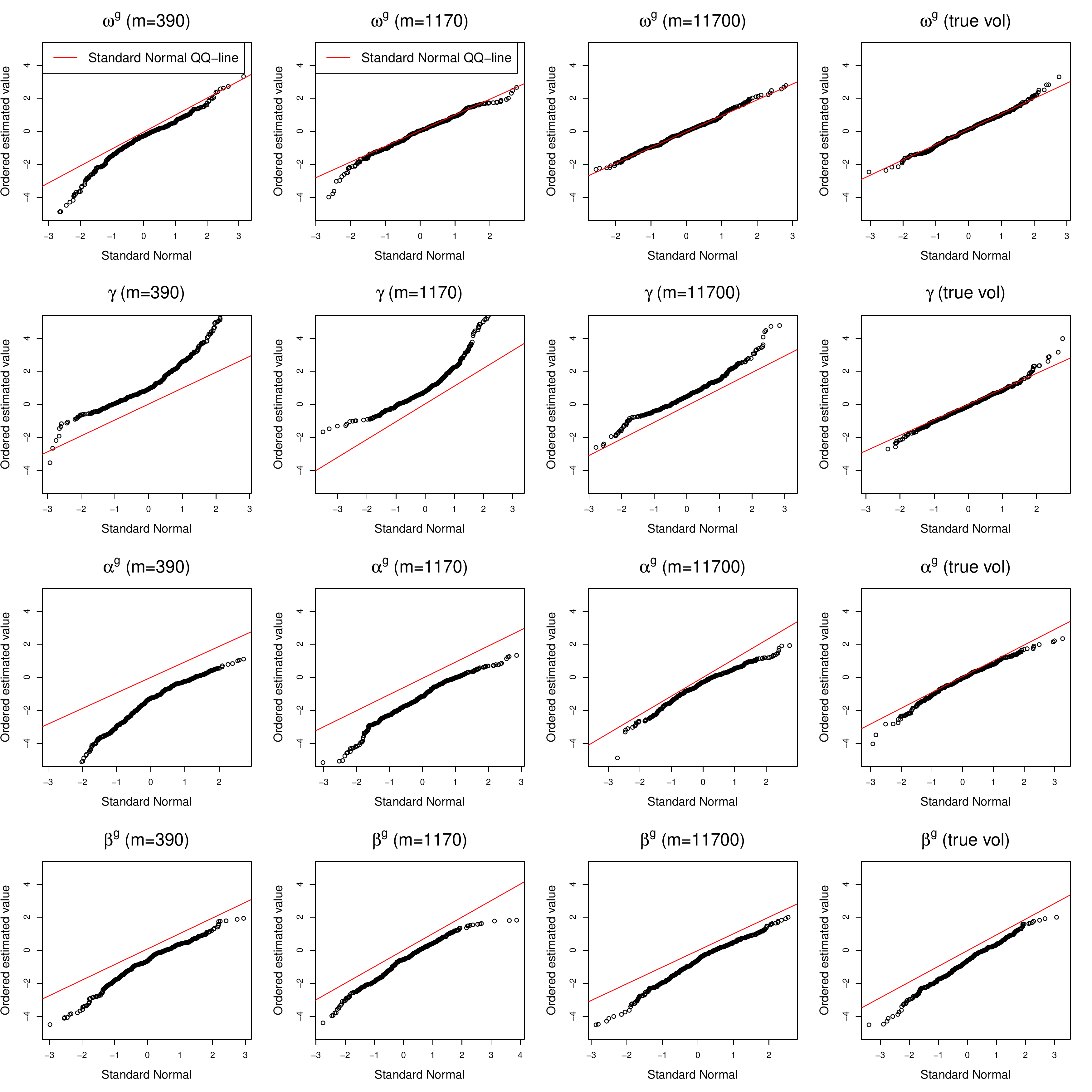}
     \caption{Standard normal quantile-quantile plots of   the Z-statistics estimates of $\omega^g$, $\gamma$, $\alpha^g$, and $\beta^g$ for $n=500$ and $m= 370, 1170, 11700$ and true volatility.
     The red real line denotes the best linear fit line, which illustrates perfect standard normal distribution.}
     \label{Figure-QQ}
\end{figure}

To check the asymptotic normality of the GARCH parameters $(\omega^g,\gamma, \alpha^g, \beta^g)$, we calculated the Z-statistics defined in Section \ref{SEC-Normality}.
Figure \ref{Figure-QQ} draws the standard normal quantile-quantile plots of   the Z-statistics estimates of $\omega^g$, $\gamma$, $\alpha^g$, and $\beta^g$ for $n=500$ and $m=390, 1170, 11700$ and true volatility.
Figure \ref{Figure-QQ} shows that, as the realized volatility closes to the true integrated volatility, the Z-statistics close to  the standard normal distribution. 
This result agrees with the theoretical conclusions in Section \ref{SEC-WLSE}. 
Thus, based on the proposed Z-statistics, we can conduct hypothesis tests using  the standard normal distribution.


%

 One of our main goals in this paper is to predict future volatility. 
 We therefore examined the out-of-sample performance of estimating the one-day-ahead GARCH volatility $h_{n+1} (\theta_0)$. 
To estimate future GARCH volatility, we  employed the proposed conditional GARCH volatility estimator $\hat{h}_{n+1} (\hat{\theta}) $, realized GARCH volatility estimator \citep{hansen2012realized, Song2018Volatlity} with only the open-to-close high-frequency observations, discrete GARCH(1,1) volatility estimator with the open-to-open log-returns, and sample variance of the open-to-open log-returns using the in-sample data.
For example, the realized GARCH volatility has the following GARCH form:
\begin{equation*} 
	h_n (\theta) = \omega + \gamma h_{n-1}(\theta) + \alpha RV_{n-1},
\end{equation*}
and the discrete GARCH(1,1) has the following GARCH form:
\begin{equation*} 
	h_n (\theta) = \omega + \gamma h_{n-1}(\theta) + \beta (X_{n-1} -X_{n-2})^2.
\end{equation*}
We then adopted the  QMLE method with the Gaussian quasi-likelihood function to estimate their GARCH parameters. 
Because the realized GARCH volatility estimator only covers the open-to-close period, we magnified the estimator by multiplying it with  
$(1 + mean[OV/RV])$ to match the magnitude, where $OV$ is the overnight return squares and $RV$ is the open-to-close realized volatility.
We call this the adjusted realized GARCH volatility. 
Finally, we also consider other estimation procedures based on Theorem \ref{prop-integratedVol} (a).
For example, we  estimate the open-to-close and close-to-open separately, as described in the first step in Section \ref{SEC-WLSE}. 
That is, without the common $\gamma$ assumption, we make inferences for $h_{n}^H(\theta)$ and $h_n^L (\theta)$ separately, using the QMLE method with the normal likelihood function. We call this the separate OGI (S-OGI) model. In contrast, 
we estimate the open-to-open conditional volatility $h_n (\theta)$ directly. 
Specifically,  Theorem \ref{prop-integratedVol} (a) shows that the conditional volatility is   
\begin{equation*} 
	h_n (\theta^g) = \omega^g + \gamma h_{n-1}(\theta) + \frac{\alpha ^g}{1-\lambda} RV_{n-1} + \frac{\beta^g}{\lambda} ( X_{n-1} - X_{\lambda+ n-2})^2.
\end{equation*}
Then we estimate the GARCH parameter  $(\omega^g,  \gamma,  \alpha ^g,  \beta^g)$ using the QMLE method with $RV +OV$ as the proxy. 
We call this the aggregated OGI (A-OGI) model. 
We note that this model can be considered as the realized GARCH model with the additional overnight innovation term.
 We measure the mean absolute errors with the one-day-ahead sample period over 500 samples as follows:
 $$
 \frac{1}{500} \sum_{i=1}^{500}| \hat{\var}_{n+1,i} - h_{n+1,i} (\theta_0)  |,
 $$
 where $ \hat{\var}_{n+1,i}$ is one of the above future volatility estimators at the $i$th sample path given the available information at time $n$. 
     \begin{table}[b] \small
\begin{center}
\caption{Mean absolute errors (MAE) for the OGI, S-OGI, A-OGI, realized GARCH, adjusted realized GARCH, GARCH, and sample variance with respect to the OGI with $n=100, 200, 500$, and $m= 390, 1170, 11700$, and true volatility.}%
\begin{tabular*}{1 \textwidth}{@{\extracolsep{\fill}}cc ccccccccc}
 \hline
 &&  \multicolumn{7}{c}{MAE$\times 10$} \\ \cline{3-9} 
$n$	&	$m$	&	OGI	&	S-OGI	&	A-OGI	&	Adj-Realized	&	Realized	&	GARCH	&	Sample	\\ \hline
100	&	390	&	0.3828 	&	0.4877 	&	0.4346 	&	0.8357 	&	1.4920 	&	0.7644 	&	1.5040 	\\
	&	1170	&	0.3612 	&	0.4500 	&	0.4224 	&	0.4574 	&	1.4287 	&	0.7644 	&	1.4415 	\\
	&	2340	&	0.3328 	&	0.4196 	&	0.4038 	&	0.3531 	&	1.3775 	&	0.7644 	&	1.3968 	\\
	&	True vol	&	0.2991 	&	0.4165 	&	0.3712 	&	0.2966 	&	1.4079 	&	0.7644 	&	1.4320 	\\
	&		&		&		&		&		&		&		&		\\
200	&	390	&	0.3090 	&	0.3561 	&	0.3356 	&	0.8186 	&	1.4986 	&	0.5170 	&	1.5027 	\\
	&	1170	&	0.2847 	&	0.3385 	&	0.3228 	&	0.4567 	&	1.4325 	&	0.5170 	&	1.4386 	\\
	&	2340	&	0.2570 	&	0.3033 	&	0.2909 	&	0.3365 	&	1.3821 	&	0.5170 	&	1.3873 	\\
	&	True vol	&	0.2353 	&	0.2881 	&	0.2674 	&	0.2855 	&	1.4086 	&	0.5170 	&	1.4163 	\\
	&		&		&		&		&		&		&		&		\\
500	&	390	&	0.2460 	&	0.2527 	&	0.2542 	&	0.7983 	&	1.5168 	&	0.4298 	&	1.5295 	\\
	&	1170	&	0.2083 	&	0.2271 	&	0.2280 	&	0.4270 	&	1.4526 	&	0.4298 	&	1.4566 	\\
	&	2340	&	0.1762 	&	0.1984 	&	0.2004 	&	0.3270 	&	1.4013 	&	0.4298 	&	1.4029 	\\
	&	True vol	&	0.1462 	&	0.1735 	&	0.1744 	&	0.2840 	&	1.4271 	&	0.4298 	&	1.4316 	\\ \hline

	\end{tabular*}
\label{Table-GARCH}
\end{center}
\end{table}

 We report the mean absolute errors for the OGI, S-OGI, A-OGI, realized GARCH, adjusted realized GARCH, GARCH, and sample variance with respect to the OGI against varying the number, $n$, of the low-frequency observations and the number, $m$, of the high-frequency observations in Table \ref{Table-GARCH}. 
In  Table \ref{Table-GARCH}, we find that the OGI models can estimate the one-day-ahead  GARCH volatility $h_{n+1} (\theta_0)$ well, but the other estimators cannot account for it well. 
 This may be because, under the OGI model, the market dynamics are explained by the open-to-close high-frequency volatility and squared close-to-open  log-returns; however, the other models ignore one of the factors. 
Compared to estimation methods for the OGI models, the WLSE yields better performance than the others. 
One possible explanation for this  is that the WLSE procedure gives more weight to the high-frequency observations, and this helps reduce the estimation errors. 
 From these results, we can conjecture that modeling appropriate overnight processes helps to not only account for market dynamics but also improve the estimation accuracy.

 \section{An empirical study}\label{SEC-empirical}

We applied the proposed OGI model to real trading high-frequency data. 
We obtained the top 5 trading volume assets (BAC,  FCX,   INTC, MSFT, MU) intra-day data from January 2010 to December 2016 from the TAQ database in the Wharton Research Data Services (WRDS) system, 1762 trading days in total.
We defined  the trading hours from 9:30 to 16:00   as the open-to-close period and the overnight period from 16:00 to the following-day 9:30 as the close-to-open period, that is, $\lambda= 6.5/24$.  
We used the log-prices and adopted the jump robust PRV estimation procedure in \eqref{eq-5.1} in the supplement document to estimate open-to-close integrated volatility.
In the empirical study, we chose the tuning parameter $c_\tau$ as  10 times the sample standard deviation of pre-averaged prices $ m^{1/8} \bar{Y}  (t_{d,k})$. 
We fixed the in-sample period as 500 days and used the rolling window scheme to estimate the parameters.



To check the relative importance of each OGI model component, we report the average proportion of jumps, and the  mean and standard deviation of the  PRV, squared overnight return, and estimated GARCH volatility from the OGI model in Table \ref{Table-summary}. 
 From Table \ref{Table-summary}, we find that the magnitude of squared overnight returns is comparable to that of PRV,  and the squared overnight returns have a greater standard deviation.
 This result leads us to conjecture that the overnight risk usually significantly affects the volatility dynamics structure.

 \begin{table}[h]
	\centering
 
\caption{Average of the jump proportion, and mean and standard deviations for pre-averaging realized volatility (PRV), overnight volatility (OV), and conditional GARCH.}%
\begin{tabular} {l ccc cccc cc}
 \hline
 & && \multicolumn{3}{c} {Mean$\times 10^4$ } &&  \multicolumn{3}{c} {SD$\times 10^4$ } \\ \cline{4-6} \cline{8-10}
Stock	&	Jump	&&	PRV	&	OV	&	GARCH	&&	PRV	&	OV	&	GARCH \\ \hline
BAC	&	0.148	&&	1.719	&	1.139	&	3.204	&&	1.577	&	3.195	&	2.571	\\
FCX	&	0.167	&&	5.423	&	3.399	&	9.080	&&	8.721	&	12.438	&	11.561	\\
INTC	&	0.158	&&	1.090	&	0.750	&	1.906	&&	0.969	&	3.675	&	1.201	\\
MSFT	&	0.141	&&	0.996	&	0.953	&	1.893	&&	1.037	&	5.254	&	1.088	\\
MU	&	0.171	&&	4.104	&	2.336	&	6.529	&&	2.990	&	11.141	&	2.732	\\ \hline

\end{tabular}
\label{Table-summary}
\end{table}

For a comparison, we calculated the OGI, S-OGI, A-OGI,  discrete GARCH(1,1), and adjusted realized GARCH volatilities defined in Section \ref{SEC-simulation} and the GJR GARCH (1,1) \citep{glosten1993relation}.
To check the performance of the ARFI-type model, we adopted the HAR-RV model \citep{corsi2009simple} and log-HAR-RV model with the bias correction \citep{demetrescu2020bias}.
For the log-HAR-RV model, we apply the HAR model on the logarithm of the realized volatility and multiply $\exp \(\hat{\sigma}^2 /2 \)$ to the forecast value, where $\hat{\sigma}^2$ is the consistent estimator of the error variance for the HAR model on the log-realized volatility. 
Then, we magnified these two estimators by multiplying $(1 + mean[OV/RV])$ to match the scale of the whole day variation, which are called ``adjusted HAR'' and   ``adjusted log-HAR,'' respectively.
To check the leverage effect, we also considered some variations of the OGI model as follows:
			\begin{equation*}	 
  h_n(\theta) = \(  \omega^g  +  \gamma h_{n-1} (\theta) + \frac{ \alpha  ^g +  I_{n-1} ^H a       }{ \lambda} \int_{n-2} ^{n-2+\lambda} \sigma_t^2 (\theta)dt  +\frac{ \beta ^g + I_{n-1} ^L b  }{ 1-\lambda } (X_{n-1} - X_{n-2+\lambda}  ) ^2 \)  ,  \\
			\end{equation*}
where $I_{n} ^H = \1_{\{ X_{n-1+\lambda} -  X_{n-1}    < c_H  \} } $, $I_{n} ^L = \1_{\{ X_{n} - X_{n-1 +\lambda} < c_L \} } $, $a$, $b$, $c_H$ and $c_L$ are additional parameters.    
To estimate the parameters, we adopted the QMLE method, which is referred to as GJR-OGI.

To measure the performance of the volatility, we used the mean squared prediction errors (MSPE) and QLIKE \citep{patton2011volatility}  as follows:
 \begin{eqnarray*}
&&MSPE= \frac{1}{n} \sum_{i=1}^n \( Vol_i - (RV_i +(X_i- X_{\lambda+i-1})^2 \)^2,\cr
&&QLIKE=  \frac{1}{n} \sum_{i=1}^n  \log Vol_i + \frac{ RV_i +(X_i- X_{\lambda+i-1})^2 }{ Vol_i} , 
 \end{eqnarray*}
where $V\!ol_i $ is one of the OGI, S-OGI, A-OGI,  GJR-OGI, GJR GARCH,  discrete GARCH, adjusted realized GARCH, adjusted HAR, and adjusted log-HAR  volatilities, and we used $RV_i +(X_i- X_{\lambda+i-1} ) ^2$ as the nonparametric daily volatility estimator.
We  predicted the one-day-ahead conditional expected volatility using the in-sample period data. 
Also, to check the significance of the difference in performances, we conducted the Diebold and Mariano (DM) test \citep{diebold1995paring} for the MSPE and QLIKE.
We compared the OGI with other models.
Table \ref{Table-rank} reports the average rank and the number of the first rank of MSPEs and QLIKEs for the nine models over the five assets.  
Table \ref{Table-full} reports the full  MSPEs and QLIKEs results, and Table \ref{DM} shows the p-values for the DM tests.
From Tables \ref{Table-rank} and \ref{Table-full}, we find that incorporating the squared overnight returns as a new innovation helps explain the market dynamics.
From Tables \ref{Table-rank}--\ref{DM}, we find that the OGI model shows the best performance overall.
This may be because  the weighted least squared estimation method helps improve the volatility prediction accuracy.


\begin{table}[h]
	\centering
 
\caption{Average rank of MSPEs and QLIKEs for the OGI, S-OGI,  A-OGI, GJR-OGI, GJR, discrete GARCH, adjusted realized GARCH, adjusted HAR,  and adjusted log-HAR. 
In the parenthesis, we report the number of the first rank among competitors.}
 
 \footnotesize{
\begin{tabular*}{1 \textwidth}{@{\extracolsep{\fill}} l ccccccccc}
 \hline
	&	OGI	&	S-OGI	&	A-OGI	&	GJR-OGI	&	GJR	&	GARCH	&	Adj-Realized	&	Adj-HAR	&	Adj-log-HAR \\ \hline
MSPE	&	2.2 (3)	&	2.8 (1)	&	4.0 (0)	&	6.2 (0)	&	9.0 (0)	&	7.2 (0)	&	5.0 (0)	&	5.2 (0) &	3.4 (1)	\\
QLIKE	&	1.6 (4)	&	3.6 (0)	&	3.8 (1)	&	3.4 (0)	&	8.4 (0)	&	8.6 (0)	&	6.0 (0)	&	4.6 (0) &	5.0 (0)	\\ \hline
\end{tabular*}
 }
\label{Table-rank}
\end{table}

\begin{table}[h]
	\centering
 
\caption{MSPEs and QLIKEs for the OGI, S-OGI,  A-OGI, GJR-OGI, GJR, discrete GARCH, adjusted realized GARCH, adjusted HAR,  and adjusted log-HAR.}%
 \scriptsize{
\begin{tabular*} {1 \textwidth}{@{\extracolsep{\fill}} ll cccccccccc}
 \hline
Stock	&		&	OGI	&	S-OGI	&	A-OGI	&	GJR-OGI	&	GJR	&	GARCH	&	Adj-Realized	&	Adj-HAR	&	Adj-log-HAR	\\ \hline
BAC	&	MSPE$\times 10^7$	&	1.194 	&	1.118 	&	1.132 	&	1.153 	&	2.110 	&	1.945 	&	1.267 	&	1.262 &	1.257	\\
	&	QLIKE	&	-7.354 	&	-7.375 	&	-7.376 	&	-7.366 	&	-7.205 	&	-7.260	&	-7.336 	&	-7.334  &	-7.340	\\
	&		&		&		&		&		&		&		&		&	 &	\\
FCX	&	MSPE$\times 10^6$	&	1.753 	&	1.894 	&	1.898 	&	2.024 	&	5.883 	&	4.212 	&	2.362 	&	2.377 &	2.275	\\
	&	QLIKE	&	-6.449 	&	-6.384 	&	-6.400 	&	-6.387 	&	-6.254 	&	-6.311 	&	-6.352 	&	-6.359  &	-6.367	\\
	&		&		&		&		&		&		&		&		&	 &	\\
INTC	&	MSPE$\times 10^7$	&	1.499 	&	1.504 	&	1.540 	&	1.586 	&	1.592 	&	1.548 	&	1.492 	&	1.495 &	1.464	\\
	&	QLIKE	&	-7.676 	&	-7.672 	&	-7.666 	&	-7.676 	&	-7.644 	&	-7.642 	&	-7.671 	&	-7.669 	&	-7.668 \\
	&		&		&		&		&		&		&		&		&	 &	\\
MSFT	&	MSPE$\times 10^7$	&	2.905 	&	2.932 	&	2.992 	&	3.099 	&	3.143 	&	2.971 	&	2.945 	&	2.957 &	2.907	\\
	&	QLIKE	&	-7.596 	&	-7.588 	&	-7.576 	&	-7.590 	&	-7.573 	&	-7.547 	&	-7.585 	&	-7.595 &	 -7.591	\\
	&		&		&		&		&		&		&		&		&	 &	\\
MU	&	MSPE$\times 10^6$	&	1.328 	&	1.354 	&	1.332 	&	1.471 	&	1.487 	&	1.438 	&	1.371 	&	1.358  &	1.356	\\
	&	QLIKE	&	-6.375 	&	-6.364 	&	-6.367 	&	-6.362 	&	-6.312 	&	-6.063 	&	-6.358 	&	-6.365  &	-6.360	\\ \hline

\end{tabular*}
}
\label{Table-full}
\end{table}

\begin{table}[h]
\centering

\caption{The p-values for the test statistic DM based on the MSPE and QLIKE for the S-OGI,  A-OGI, GJR-OGI, GJR, discrete GARCH, adjusted realized GARCH, adjusted HAR,  and adjusted log-HAR with respect to the OGI.}
\scriptsize{
\begin{tabular*}{1 \textwidth}{@{\extracolsep{\fill}}l l l l c c c c c c c c}
\hline
Stock &&  && 	S-OGI	&	A-OGI	&	GJR-OGI	&	GJR	&	GARCH	&	Adj-Realized	&	Adj-HAR	&	Adj-log-HAR \\ \hline
BAC &&	 $\text{MSPE}$   &&	 0.999     &  0.823 	&	 0.730  &	 0.000  	&  0.000  	&  0.245  	&  0.265  &  0.283    	\\
 &&	 $\text{QLIKE}$   &&	 0.999    &  0.999 	&	 0.975 &	 0.000 	& 0.000 	&  0.072 	&  0.050  &  0.148  	\\
&& && \\

FCX &&	 $\text{MSPE}$   &&	 0.011    &  0.035 	&	 0.059 &	 0.000  	&  0.000 	& 0.000  	&  0.000  &  0.000    	\\
 &&	 $\text{QLIKE}$   &&	 0.000    &  0.001 	&	 0.021 &	  0.000 	&  0.000 	&  0.000 	&  0.002  &  0.004  	\\
&& && \\

INTC &&	 $\text{MSPE}$   &&	 0.328     &  0.019 	&	 0.145  &	 0.054  	&  0.159  	&  0.563  	&  0.542  &  0.817    	\\
 &&	 $\text{QLIKE}$   &&	 0.371    &  0.280 	&	 0.479 &	  0.010 	&   0.013 	&   0.379 	&  0.238  &  0.311  	\\
&& && \\

MSFT &&	 $\text{MSPE}$   &&	 0.015     &  0.001 	&	 0.020  &	 0.004  	&  0.060  	&  0.112  	&  0.170  &  0.463    	\\
 &&	 $\text{QLIKE}$   &&	  0.164    &  0.036  	&	 0.343 &	  0.158 	&  0.000 	&  0.276 	&  0.469  &  0.395  	\\
&& && \\

MU &&		 $\text{MSPE}$   &&	0.110     &  0.105 	&	 0.069  &	 0.000  	&  0.000  	&  0.000 	&  0.004  &  0.003    	\\
 &&	 $\text{QLIKE}$   &&	 0.191    &  0.185 	&	 0.063 &	 0.036 	&  0.122 	&  0.029 	&   0.091  &  0.100  	\\ \hline
\end{tabular*}
}
\label{DM}
\end{table}

To check the benefit of using the overnight information, we employed a utility based framework \citep{bollerslev2018modeling, demetrescu2020bias, fleming2003economic}.
Specifically, we assume that there is a risk-averse person who invests part of his wealth $x_i \in [0, 1]$ in specified asset and holds the remaining part $1-x_i$ at time $i-1$.
The mean-variance utility function is defined as
\begin{equation*}
EU(R_i)=\E\left[x_i R_i \middle | \FF_{i-1} \right] - \dfrac{\xi}{2}\Var \left[x_i R_i \middle | \FF_{i-1}\right],
\end{equation*} 
where $\xi>0$ is the risk aversion coefficient and $R_i$ is the daily log-return.
Then, the optimal allocation has the following form:
\begin{equation*}
x_i=\dfrac{1}{\xi}\dfrac{\E\left[ R_i \middle | \FF_{i-1} \right]}{\Var \left[R_i \middle | \FF_{i-1}\right]},
\end{equation*} 
where $\E\left[ R_i \middle | \FF_{i-1} \right]$ and $\Var \left[R_i \middle | \FF_{i-1}\right]$ are replaced by their forecasts.
Since it is difficult to predict the expectation $\E\left[ R_i \middle | \FF_{i-1} \right]$  in practice, we replace it with the previous day's log-return based on the martingale assumption on the daily return process.
Also, the variance $\Var \left[R_i \middle | \FF_{i-1}\right]$ is replaced by one of the OGI, S-OGI, A-OGI,  GJR-OGI, GJR, discrete GARCH, realized GARCH, HAR, and log-HAR  volatilities.
For the OGI, S-OGI, A-OGI,  GJR-OGI, GJR, and GARCH, we used the open-to-open returns for $R_i$, while the open-to-close returns were used for the realized GARCH, HAR, and log-HAR. 
Then, we obtained $\hat{x}_i$ and set
$ x_i^{\ast}   = \hat{x}_i \1 _{\{\hat{x}_i \in [0, 1]\}} + \1 _{\{\hat{x}_i>1\}}$
to make the investment feasible and prevent the short-sellings.
Finally, the resulting returns $R_i^{\ast}$=$x_i^{\ast} R_i$ were used to measure the economic performances.
Specifically, with the mean $\bar{R}^{\ast}$ and standard deviation $\bar{S}^{\ast}$ of the returns $R_i^{\ast}$, we calculated the Sharpe ratio $SR^{\ast} = \bar{R}^{\ast} / \bar{S}^{\ast}$ and expected utility $EU^{\ast}=\bar{R}^{\ast} -\dfrac{\xi}{2}\( \bar{S}^{\ast}\)^2$.
Table \ref{Sharpe} reports the Sharpe ratios and expected utilities  for the nine models for $\xi=2.5, 5$ over the five assets.  
As seen in Table \ref{Sharpe}, the models with the open-to-open information show better performance than  other models.
This may indicate that considering overnight period  helps obtain the additional economic gains.
When comparing the models with the open-to-open information, the OGI-based models do not significantly outperform the GJR and GARCH models. 
This may be because the future return estimator often has the huge errors in practice. 
From this result, we can conjecture that the overnight period is significant in terms of investing strategy.

\begin{table}[h]
	\centering
 
\caption{Sharpe ratios ($SR^{\ast}$) and expected utilities ($EU^{\ast}$) for the OGI, S-OGI,  A-OGI, GJR-OGI, GJR, discrete GARCH, realized GARCH, HAR,  and log-HAR for $\xi=2.5, 5$ and five assets.}%
 \scriptsize{
\begin{tabular*} {1 \textwidth}{@{\extracolsep{\fill}} lll cccccccccc}
 \hline
Risk aversion & Stock	&		&	OGI	&	S-OGI	&	A-OGI	&	GJR-OGI	&	GJR	&	GARCH	&	Realized	&	HAR	&	Log-HAR	\\ \hline
$\xi$=2.5 & BAC	&	$SR^{\ast}$ $\times 10^2$	&	3.974 	&	3.952 	&	3.944	&	3.911 	&	3.936 	&	3.882 	&	-5.173 	&  -5.167  &	 -5.166	\\
  &	   &	$EU^{\ast}$ $\times 10^4$	&	3.116 	&	3.086	&	3.073 	&	3.027 	&	3.062 	&	2.985	&	-6.864 	&	-6.854  &	-6.855	\\
  & 	   &		&		&		&		&		&		&		&		&	 &	\\
  & FCX	&	$SR^{\ast}$ $\times 10^2$	&	-2.891 	&	-2.993 	&	-3.004 	&	-3.002 	&	-2.894 	&	-2.869 	& -5.008 	&	-5.002 &	 -4.996	\\
  &	   & $EU^{\ast}$ $\times 10^3$	&	-1.333 	&	-1.350 	&	-1.352 	&	-1.357 	&	-1.318 	&	-1.312 	&	-1.386 	&	-1.386  &	-1.386	\\
  &   &		&		&		&		&		&		&		&		&	 &	\\
  &INTC	&	$SR^{\ast}$ $\times 10^2$	&	2.302 	&	2.301 	&	2.305 	&	2.307 	&	2.323 	&	2.306 	&	5.368 	&	5.370 &	5.363	\\
  &  	   &	$EU^{\ast}$	$\times 10^4$ &	1.060	&	1.059 	&	1.063 	&	1.064 	&	1.079 	&	1.064 	&	3.441 	&	3.443 	&	3.437 \\
  &   &		&		&		&		&		&		&		&		&	 &	\\
  & MSFT	&	$SR^{\ast}$ $\times 10^2$	&	2.857 	&	2.909 	&	2.925 	&	2.907 	&	2.848 	&	2.812 	&	2.789 	&	2.769  &	2.787	\\
  &      &	$EU^{\ast}$ $\times 10^4$ 	&	1.588 	&	1.639 	&	1.654 	&	1.637 	&	1.579 	&	1.543 	&	1.375 	&	1.360  &	 1.374	\\
	&   &		&		&		&		&		&		&		&		&	 &	\\
  &  MU	&	$SR^{\ast}$ $\times 10^2$	&	5.875 	&	5.932 	&	5.878 	&	5.839 	&	5.642 	&	5.651 	&	0.007 	& -0.029  &	 -0.038	\\
  &     &	$EU^{\ast}$	$\times 10^4$ &	6.691 	&	6.803 	&	6.695 	&	6.615 	&	6.230 	&	6.250 	&	-2.875 	&	-2.944  &	-2.955	\\ \hline
$\xi$=5 & BAC	&	$SR^{\ast}$ $\times 10^2$	&	3.726	 &	3.813 	&	3.822 	&	3.724 	&	3.703 	&	3.646 	&	-5.018 	&  -5.029  &	 -5.027	\\
  &	   &	$EU^{\ast}$ $\times 10^4$	&	0.344 	 &	0.455	&	0.466 	&	0.336 	&	0.348 	&	0.252	&	-8.023	&	-8.025  &	-8.032	\\
  & 	   &		&		&		&		&		&		&		&		&	 &	\\
  & FCX	&	$SR^{\ast}$ $\times 10^2$	&	-2.937 	&	-2.921 	&	-2.998 	&	-2.792 	&	-2.891 	&	-2.842 	& -5.108 	&	-5.182 &	 -5.182	\\
  &	   & $EU^{\ast}$ $\times 10^3$	&	-1.972 	&	-1.940 	&	-1.956 	&	-1.924 	&	-1.898 	&	-1.893 	&	-1.798 	&	-1.818  &	-1.817	\\
  &   &		&		&		&		&		&		&		&		&	 &	\\
  &INTC	&	$SR^{\ast}$ $\times 10^2$	&	2.483 	&	2.424 	&	2.436 	&	2.466 	&	2.492 	&	2.431 	&	5.199 	&	5.121 &	5.167	\\
  &  	   &	$EU^{\ast}$	$\times 10^4$ &	0.177	   &	0.120 	&	0.131 	&	0.161 	&	0.183 	&	0.127 	&	2.535 	&	2.473 	&	2.510 \\
  &   &		&		&		&		&		&		&		&		&	 &	\\
  & MSFT	&	$SR^{\ast}$ $\times 10^2$	&	2.550 	&	2.680 	&	2.733 	&	2.665 	&	2.477 	&	2.426 	&	2.492 	&	2.502  &	2.498	\\
  &      &	$EU^{\ast}$ $\times 10^4$ 	&	0.194 	&	0.317 	&	0.366 	&	0.297 	&	0.134 	&	0.089 	&	0.490 	&	0.497  &	 0.494	\\
	&   &		&		&		&		&		&		&		&		&	 &	\\
  &  MU	&	$SR^{\ast}$ $\times 10^2$	&	5.372 	&	5.351 	&	5.314 	&	5.274 	&	5.372 	&	5.162 	&	0.172 	&  0.178  &	 0.169	\\
  &     &	$EU^{\ast}$	$\times 10^4$ &	1.264 	&	1.219 	&	1.143 	&	1.093 	&	1.283 	&	0.918 	&	-5.260 	&	-5.259  &	-5.277	\\ \hline
  
\end{tabular*}
}
\label{Sharpe}
\end{table}

To check the volatility persistence of the nonparametric volatility, we study  regression residuals between the nonparametric volatility and estimated conditional volatilities. 
Specifically, we fitted the following linear model 
$$
  RV_i +(X_i- X_{\lambda+i-1})^2 = a +b\times Vol_i +e_i,
$$
where $Vol_i $ is one of the predicted volatilities OGI, S-OGI,  A-OGI, GJR-OGI, GJR, discrete GARCH, adjusted realized GARCH, adjusted HAR,  and adjusted log-HAR. 
Then, we calculated the regression residuals, $\hat{\epsilon}_i$, for each model and checked their auto-correlations over lag $L=1, \ldots, 30$. 
Table \ref{Table-rank2} reports the average rank and the number of the first rank for the first and max absolute auto-correlations from the nine models. 
In the supplement document, we draw the auto-correlation function (ACF) of the regression residuals for each model and asset (see Figure \ref{Figure-ACF}). 
From Table \ref{Table-rank2} and Figure \ref{Figure-ACF}, we find that the OGI, S-OGI, A-OGI, and GJR-OGI models show better performance than  other estimators overall.
That is, the OGI-based models can explain the market dynamics in the volatility time series.

 \begin{table}[h]
	\centering
 
\caption{Average ranks in order from the smallest to the biggest for the first and max absolute auto-correlations over lag $L=1, \ldots, 30$ for the OGI, S-OGI,  A-OGI, GJR-OGI, GJR, discrete GARCH, adjusted realized GARCH, adjusted HAR,  and adjusted log-HAR. 
In the parenthesis, we report the number of the first rank among competitors.}
 
 \footnotesize{
\begin{tabular*}{1 \textwidth}{@{\extracolsep{\fill}} l ccccccccc}
 \hline
	&	OGI	&	S-OGI	&	A-OGI	&	GJR-OGI	&	GJR	&	GARCH	&	Adj-Realized	&	Adj-HAR	&	Adj-log-HAR \\ \hline
First	&	1.8 (3)	&	3.8 (0)	&	4.0 (0)	&	3.0 (1)	&	4.6 (1)	&	6.4 (0)	&	6.8 (0)	&	6.6 (0) &	8.0 (0)	\\
Max	&	2.8 (1)	&	3.0 (2)	&	3.8 (0)	&	5.6 (0)	&	6.8 (0)	&	6.4 (0)	&	5.4 (0)	&	4.8 (1) &	6.4 (1)	\\ \hline
\end{tabular*}
 }
\label{Table-rank2}
\end{table}

 We examined the performance of the proposed method in measuring  one-day-ahead VaR.
To evaluate VaR,  we first predicted the one-day-ahead conditional expected volatility by  the OGI, S-OGI,  A-OGI, GJR-OGI, GJR, discrete GARCH, adjusted realized GARCH, adjusted HAR, and adjusted log-HAR using the in-sample period data. 
We then calculated the quantiles by historical standardized daily returns. 
Specifically, we standardized the in-sample daily returns by the fitted conditional volatilities.  
Then, we calculated the sample quantiles for 0.01, 0.02, 0.05, 0.1, 0.2 and with the sample quantile estimates and  predicted volatility, we obtained the one-day ahead VaR values. 
We fixed the in-sample period as 500 days and used the rolling window scheme.

To backtest the estimated VaR, we conducted  the likelihood ratio unconditional coverage (LRuc) test \citep{kupiec1995techniques}, the likelihood ratio conditional coverage (LRcc) \citet{christoffersen1998evaluating}, and  the dynamic quantile (DQ) test with lag 4 \citep{engle2004caviar}.
Table \ref{Table-rank3} reports the number of cases where p-value is bigger than $0.05$  for the five assets and  $q_0 = 0.01, 0.02, 0.05, 0.1, 0.2$ based on the LRuc, LRcc and DQ tests.
In the supplement document, we draw the scatterplots for the p-values of LRuc, LRcc, and DQ tests for the nine models with $q_0= 0.01, 0.02, 0.05, 0.1, 0.2$ (see Figure \ref{Figure-VaR}). 
As seen in Table \ref{Table-rank3} and  Figure \ref{Figure-VaR}, the OGI model shows the best performance for all hypothesis tests.
This result shows that  the overnight risk is important to account for whole-day market dynamics, and the OGI process can account for market dynamics by utilizing the overnight risk information.
In contrast, other OGI-based models show relatively worse performance.
This finding prompts us to speculate that it may help improve estimation accuracy by estimating the open-to-close and close-to-open separately with the weighted least squared estimation method under the common $\gamma$ condition.

 \begin{table}[h]
	\centering
 
\caption{Number of cases where p-value is bigger than $0.05$ for the OGI, S-OGI,  A-OGI, GJR-OGI, GJR, discrete GARCH, adjusted realized GARCH, adjusted HAR,  and adjusted log-HAR for the 5 assets and  $q_0 = 0.01, 0.02, 0.05, 0.1, 0.2$ based on the LRuc, LRcc and DQ tests.}
 
 \scriptsize{
\begin{tabular*}{1 \textwidth}{@{\extracolsep{\fill}} l ccccccccc}
 \hline
	&	OGI	&	S-OGI	&	A-OGI	&	GJR-OGI	&	GJR	&	GARCH	&	Adj-Realized	&	Adj-HAR	&	Adj-log-HAR \\ \hline
LRuc	& 25	&	17	&	20	&	18	&	22	&	21	&	21	&	21  &	18	\\
LRcc	& 25	&	22	&	20	&	19	&	20	&	21	&	21	&	21  &	16	\\
DQ	   & 18	&	7  &	6	&	10	&	18	&	10	&	6	&	11  &	0	\\ \hline
\end{tabular*}
 }
\label{Table-rank3}
\end{table}

\section{Conclusion}\label{SEC-6}

In this paper, we introduce the diffusion process, which can explain the whole-day volatility dynamics. 
Specifically, the proposed OGI model can account for the different dynamic structures for the open-to-close and close-to-open periods. 
To do this, we introduced the weighted QMLE procedure and showed its asymptotic properties. 
In the empirical study, we found the benefit of incorporating the overnight information. 
The models with overnight innovation term perform better than other models for the prediction of daily volatility, utility based analysis, and volatility persistence analysis.
It suggests that  incorporating the overnight information helps account for the dynamic structure of the daily total variation. 
On the other hand, the OGI model outperforms other OGI-based models in terms of the prediction of daily volatility, analysis of volatility persistence, and one-day-ahead VaR measurement.   
It reveals that  the weighted least squared estimation method with the common $\gamma$ condition helps obtain better estimation accuracy. 

In practice, we often observe zero returns \citep{francq2021volatility}. 
However, the proposed diffusion process cannot account for the zero return phenomena. 
Thus, it is an interesting future study to develop a diffusion process, which can account for the zero returns, and to investigate its effect.

%
%
%


\bibliography{myReferences}
\end{spacing}
\end{document}


\maketitle
\begin{spacing}{1.9}
\appendix
\counterwithin{figure}{section}
\counterwithin{table}{section}

\section{Simulation setup}

We chose the drift $\mu_t=0$, and   $(\omega_{H1,0}, \omega_{H2,0}, \omega_{L,0}, \gamma_{H,0}, \gamma_{L,0}, \alpha_{H,0}, \alpha_{L,0}, \beta_{H,0}, \beta_{L,0}, \nu_{H,0}, \nu_{L,0}) = (0.02, 0.01,$ $0.01, 0.6, 0.6, 0.4, 0.1, 0.2,0.1,0.4,0.2)$. 
In this case, the target GARCH parameter is $(\omega_{H,0}^g, \omega_{L,0}^g, \gamma_0, \alpha_{H,0}^g,$ $\alpha_{L,0}^g,\beta_{H,0}^g) = (0.067, 0.063, 0.36, 0.21, 0.202, 0.128, 0.096) $. 
 For the open-to-close period, we generated the noisy observations as follows:
 $$
 	Y_{t_{d,j}}= X_{t_{d,j}} + \epsilon_{t_{d,j}}, \quad \text{for } d=1,\ldots, n, j=1, \ldots, m-1,
 $$
 where $m$ is the number of high-frequency observations for the open-to-close period, and $\epsilon_{t_{d,j}}$'s are generated from i.i.d. normal distributions with mean zero and  standard deviation $0.01 \sqrt{\int_{d-1}^d \sigma_t ^2 (\theta) dt}$.
 To generate the true process, we chose $m^{all} = 43,200$, which equals the number of every 2 seconds in a one-day period.
 We varied $n$ from 100 to 500 and $m$ from  390 to 11,700, which correspond to the numbers of 1 minute  and every 2 seconds in the open-to-close period, respectively. 
 We treated $Y_{t_{d,j}}, j=1, \ldots, m-1$ as the high-frequency observations 
 and the open and close prices $X_{t_{d,0}}$ and $X_{t_{d,m}}$ as the observed log-prices.  
 To estimate the integrated volatility for the open-to-close period, we employed the jump adjusted pre-averaging realized volatility estimator   \citep{ait2016increased, jacod2009microstructure} as follows:
 \begin{equation}\label{eq-5.1} 
RV_d  =\frac{1} {  \psi K} \sum_{k=1}^{m-K+1} \left \{   \bar{ Y} ^2 (t_{d,k})   - \frac{1}{2} \,  \hat{ Y} ^2 (t_{d,k})    \right \}  \1_{ \{ | \bar{Y}  (t_{d,k}) |  \leq \tau_m \} },
\end{equation}
where
\begin{eqnarray*}
&& \bar{Y}  (t_{d,k}) = \sum_{l=1}^{K-1} g \( \frac{l}{K}\) \(   Y _{t_{d, k+l}}  - Y_{t_{d, k+l-1}} \), \quad  \psi =\int_{0}^{1} g(t)^2 dt, \cr
&&  \hat{ Y} ^2 (t_{d,k})  =    \sum_{l=1}^{K} \left \{ g\(\frac{l}{K}\)-g\(\frac{l-1}{K}\) \right \}^2 \( Y_{t_{d, k+l-1}}-Y_{t_{d,k+l-2}} \)^2,   
\end{eqnarray*}
we take the weight function $g(x) = x \wedge (1-x)$  and the bandwidth size $K= \lfloor m^{1/2} \rfloor$, $\1_{\{ \cdot\} }$ is an indicator function,  and $\tau_m= c_\tau m^{- 0.235}$ is a truncation level for the constant  $c_\tau$.
 We chose $c_\tau$ as  three times the sample standard deviation of the pre-averaged prices $ m^{1/8} \bar{Y}  (t_{d,k})$.

\section{Miscellaneous materials}\label{materials}

  \begin{figure} 
  \centering
    \includegraphics[width=1\textwidth]{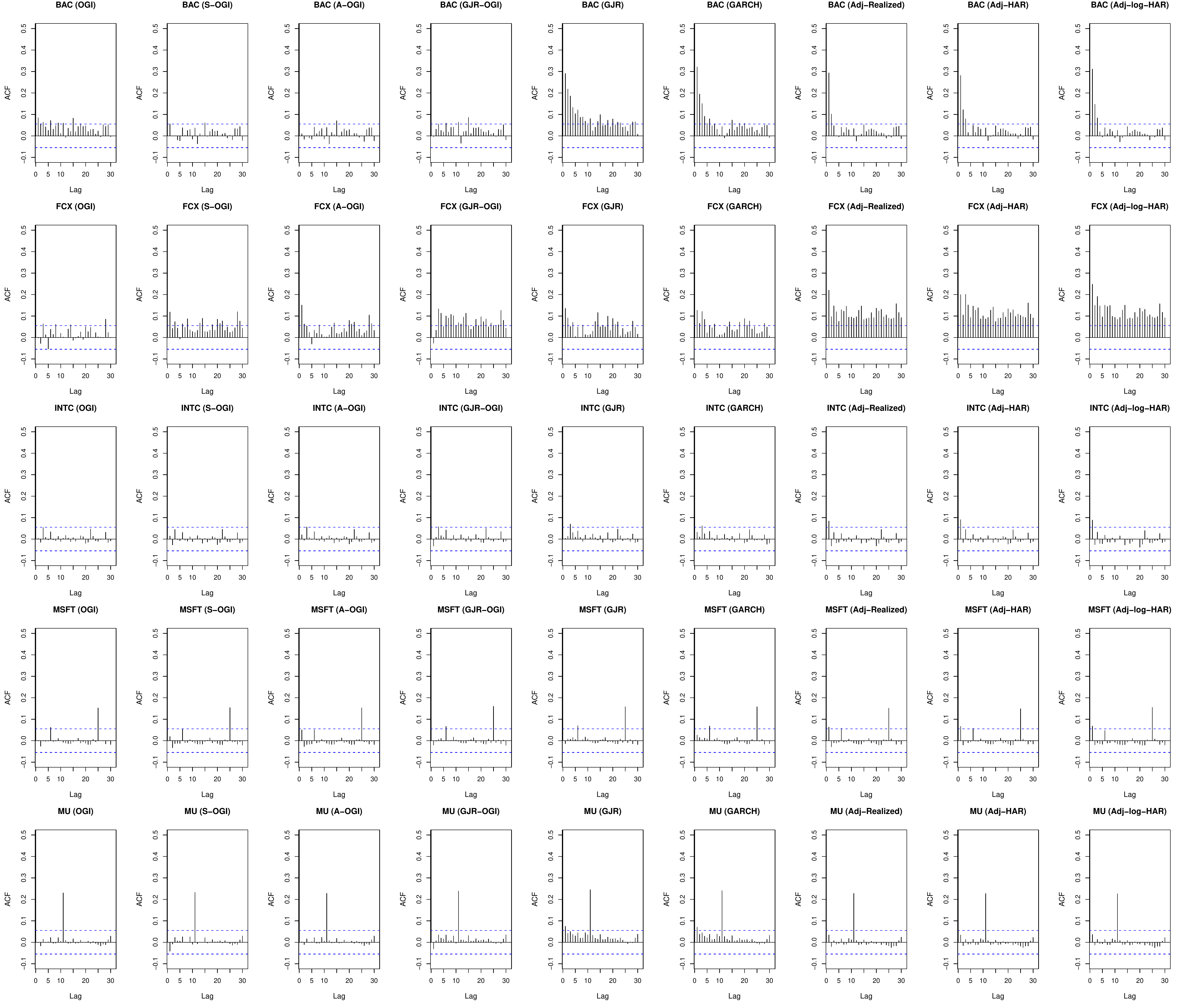}
     \caption{ACF plots for the regression residuals between the nonparametric volatility and estimated volatility, such as  the OGI, S-OGI,  A-OGI, GJR-OGI, GJR, discrete GARCH, adjusted realized GARCH, adjusted HAR, and adjusted log-HAR. }
     \label{Figure-ACF}
\end{figure}

  \begin{figure} 
  \centering
    \includegraphics[width=1\textwidth]{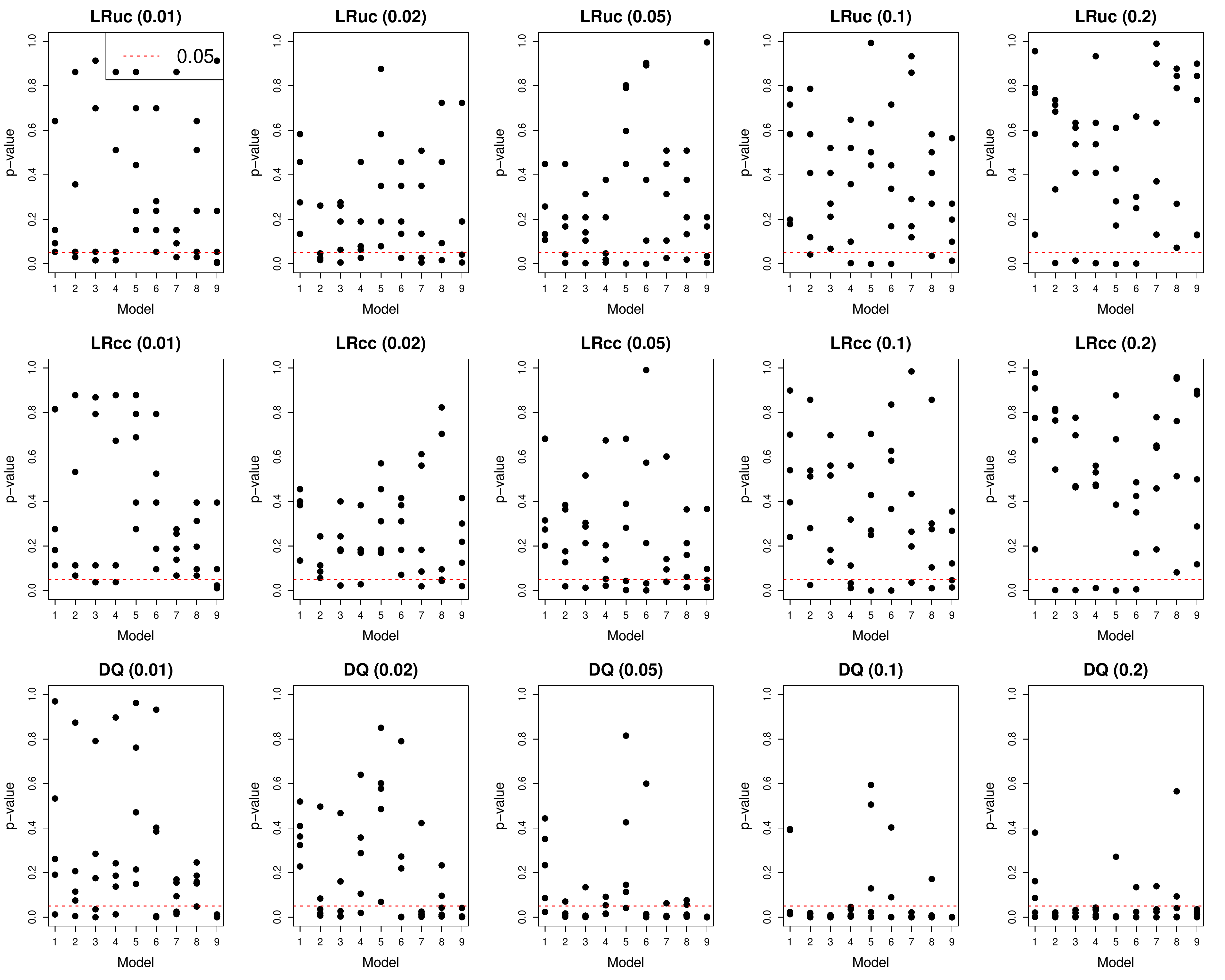}
     \caption{Scatterplots for the p-values of LRuc, LRcc, and DQ tests with $q_0=$ 0.01, 0.02, 0.05, 0.1, and 0.2. 
    Note that  the OGI (1), S-OGI (2),  A-OGI (3), GJR-OGI (4),  GJR (5), discrete GARCH (6),  adjusted realized GARCH (7), adjusted HAR (8), and adjusted log-HAR (9). }
     \label{Figure-VaR}
\end{figure}

\newpage

 \section{Proofs} \label{SEC-proof}

  \subsection{Proof of Theorem \ref{prop-integratedVol}}\label{SEC-relationship}
 
 Let  $P_t = \int_{0}^t \sigma_t (\theta) dB_t$.
  Theorem \ref{prop-integratedVol} is an immediate consequence of Theorem \ref{prop-integratedVol-apendix} (a) below.  
  
 \begin{theorem} 	\label{prop-integratedVol-apendix}
 For the OGI model, the integrated volatilities have the following structure.
	\begin{enumerate}
		\item[(a)] For $0<\alpha_H<1$ and $ n \in \mathbb{N}$,   we have 
			\begin{equation*}
			\int_{n-1} ^{ n-1 +\lambda}  \sigma_{t}^2(\theta)  dt = \lambda h_n^H (\theta)+D_n^H \quad a.s.,
			\end{equation*}
		where
			\begin{equation*}	 
			h_n ^H (\theta) =  \omega_H^g + \gamma h_{n-1}^H (\theta) + \alpha_H^g \lambda^{-1} \int_{n-2}^{ n-2+\lambda} \sigma_t ^2 (\theta) dt + \beta_{H}^g (1-\lambda)^{-1} ( P_{n-1} - P_{n-2 +\lambda} )^2, 
			\end{equation*}
  \begin{eqnarray*}
 	&& \varrho_{H1} = \alpha_H ^{-1} ( e^{\alpha_H  } -1) , \quad  \varrho_{H2} = \alpha_H ^{-2} ( e^{\alpha_H } -1 - \alpha_H  ) ,   \cr
 	&&\varrho_{H3} = \alpha_H ^{-3} ( e^{\alpha_H } -1 - \alpha_H  -   \alpha_H  ^2/2  ),\quad	\varrho_H = 2 \gamma_H   \varrho_{H3 }   + \varrho_{H1}  - \varrho_{H2}  , \cr
	&& \omega_H ^g =(1-\gamma) \[ 2  \omega_{H1}  \varrho_{H3} -   \omega_{H2} \varrho_{H2}  + \nu_H   \{    \varrho_{H2} - 2 \varrho_{H3} \}     \]  + \gamma_L (\omega_{H1} - \omega_{H2})  \varrho_{H}+ \omega_L  \varrho_H  , \cr
		&&  \gamma = \gamma_H \gamma_L,   \quad   \alpha_H^g = \varrho_H \gamma_L \alpha_H, \quad   \beta_{H}^g =\varrho_H  \beta_L  + \beta_H (\varrho_{H2} - 2 \varrho_{H3})  ,
 \end{eqnarray*}
		and
			\begin{equation*}
			D_n ^H=2 \nu_H \alpha_H^{-2} \int_{n-1}^{ n -1 +\lambda }   \{  \alpha_H \lambda^{-1} ( \lambda +n-1 -t-   \lambda \alpha_H   ^{-1} ) e^{\lambda^{-1}\alpha_H (\lambda +n -1 -t) } +1 \} Z_t^H dW_t
			\end{equation*}
		is a martingale difference.

			\item[(b)] For   $0<\beta_L<1$,  and $ n \in \mathbb{N}$,   we have 
			\begin{equation*} 
			\int_{ n-1 + \lambda} ^{n}  \sigma_{t}^2(\theta)  dt = (1-\lambda) h_n^L (\theta)+D_n^L \quad a.s.,
			\end{equation*}
		where
			\begin{equation*}	 
			h_n ^L (\theta) =  \omega_L^g + \gamma h_{n-1}^L (\theta) + \alpha_L^g \lambda^{-1} \int_{n-2}^{ n-2+ \lambda} \sigma_t ^2 (\theta) dt + \beta_{L}^g  (1-\lambda)^{-1}( P_{n-1} - P_{ n-2+\lambda} )^2, 
			\end{equation*}
  \begin{eqnarray*}
 	&& \varrho_{L1} = \beta_L ^{-1} ( e^{\beta_L  } -1) , \quad  \varrho_{L2} = \beta_L ^{-2} ( e^{\beta_L  } -1 - \beta_L   ) ,   \quad  	\varrho_L = (\gamma_L -1) \varrho_{L2 }   + \varrho_{L1} , \cr	
	&& \omega_L ^g =   (1-\gamma) \[ \omega _L \varrho_{L2} + \nu_L (\varrho_{L2} - 2\varrho_{L3} )  \]   +   ( \omega_{H1} - \omega_{H2}  + \gamma_H \omega _L) \varrho_L + \varrho_L  \alpha_H \omega_H^g  \cr
&& + \alpha_L ( \varrho_{L2} - 2\varrho_{L3})   \omega_H^g, \quad   \alpha_L^g= \varrho_L\alpha_H  (  \gamma +\alpha_H ^g) +\alpha_L ( \varrho_{L2} -2 \varrho_{L3})  ( \gamma + \alpha_H^g)   \cr
	&&  \beta_{L}^g =\varrho_L (\gamma_H  \beta_L  +\alpha_H \beta_H^g) + \alpha_L ( \varrho_{L2} - 2\varrho_{L3})   \beta_H^g  ,   
 \end{eqnarray*}
		and
			\begin{eqnarray*}
			D_n ^L  &=& 2 \int_{ n-1 + \lambda}^{n}   ( e^{\beta_L (1-\lambda) ^{-1} (n-t) } -1 )(P_{t} - P_{\lambda+n-1} ) \sigma_t (\theta) dB_t  \cr
				&& +  	2 \nu_L \beta_L^{-2} \int_{n-1+\lambda} ^{n} \[ \frac{\beta_L}{1- \lambda} \{ n-t - (1-\lambda) \beta_L^{-1} \} e^{\beta_L (1-\lambda)^{-1} (n-t) } +1 \]   Z_t^L dW_t  \cr
				&& +  ( \varrho_L  \beta_H  + \alpha_L (\varrho_{L2} -2 \varrho_{L3}) (1-\lambda) \lambda ^{-1}   D_n^H
			\end{eqnarray*}
		is a martingale difference.
		
\item [(c)]		For $0<\beta_H<1$,  $0<\beta_L<1$, and $ n \in \mathbb{N}$, we have 
\begin{equation*}
	\int_{n-1}^{n}  \sigma^2_t (\theta) dt = h_n (\theta) + D_n \quad \text{a.s.}, 
\end{equation*}
where $D_n =   D_n^H  +D_n^L $, 
		\begin{equation} \label{eq3-con-GARCH} 
		h_n(\theta) = \omega^g + \gamma h_{n-1} (\theta) + \alpha  ^g \lambda^{-1} \int_{n-2} ^{n-2+\lambda} \sigma_t^2 (\theta)dt  +\beta ^g (1-\lambda)^{-1} (P_{n-1} - P_{n-2+\lambda} ) ^2 , 
		\end{equation}
\begin{eqnarray*}
		 \omega^g = \lambda \omega_H^g + (1-\lambda) \omega_L ^g , \quad \alpha ^g = \lambda   \alpha_H^g + (1-\lambda) \gamma \alpha_L^g  ,  \quad  \beta  ^g = \lambda  \beta_{H}^g  +  (1-\lambda)\beta_{L}^g.
\end{eqnarray*}

		\item[(d)] For $0<\alpha_H<1$,  $0<\beta_L<1$, and $ n \in \mathbb{N}$, we have 
			\begin{eqnarray*}
			&& E \left[ \int_{n-1}^{n-1+\lambda}  \sigma^2_t (\theta) dt \Bigg| \mathcal{F}_{n-1} \right] = \lambda h_n^H(\theta) \quad a.s., \\ 
			&& E \left[ \int_{ n-1+\lambda}^{n}  \sigma^2_t (\theta) dt \Bigg| \mathcal{F}_{n-1} \right] = (1-\lambda) h_n^L(\theta) \quad a.s., \\ 
			&& E \left[ \int_{n-1}^{n}  \sigma^2_t (\theta) dt \Bigg| \mathcal{F}_{n-1} \right] = h_n (\theta) \quad a.s. \nonumber
			\end{eqnarray*}
		
	\end{enumerate}
\end{theorem}
 
 \textbf{Proof of Theorem \ref{prop-integratedVol-apendix}.}
Consider $(a)$ and $(b)$. 
 By It\^o's lemma, we obtain 
 \begin{eqnarray*}
 	R_H (k) &=& \int_{n-1} ^{\lambda + n-1} \frac{(\lambda +n-1-t)^k }{k!} \sigma_{t}^2(\theta)  dt  \cr
 		&=& \omega_{H1}  2  \lambda^{-2} \frac{\lambda^{k+3}}{(k+3)!} -\omega_{H2}  \lambda^{-1} \frac{\lambda ^{k+2} }{ (k+2) !}  \cr
 		&& + \sigma_{n-1}^2(\theta) \lambda^{-2}   \left \{  \gamma_H 2   \frac{\lambda^{k+3}}{(k+3)!}  -  \lambda  \frac{\lambda ^{k+2} }{ (k+2) !} + \lambda ^2 \frac{\lambda ^{k+1} }{ (k+1) !} \right \}  \cr
 		&& +  \frac{  \beta_H  }{  \lambda ^2 (1-\lambda)}   \left \{  \lambda \frac{\lambda^{k+2} }{ (k+2) ! }  - 2 \frac{\lambda^{k+3} }{ (k+3)! }   \right \}   \sum_{j=1}^\infty \gamma ^{j-1}   \( \int_{n-1+\lambda-j } ^{n-j} \sigma_s  (\theta) dB_s  \)^2  \cr
 			&&+ \nu_H \lambda^{-2} \left \{  \lambda \frac{\lambda^{k+2} }{ (k+2) ! }  - 2 \frac{\lambda^{k+3} }{ (k+3)! }   \right \} \cr
 			&& + 2 \lambda^{-2} \nu_H   \int_{n-1}^{\lambda + n -1} \frac{(\lambda +n -1 -s) ^{k+2} (k+1) }{(k+2)!} (W_s- W_{n-1} ) dW_s \cr
 			&&+ \alpha_H \lambda^{-1} R_H (k+1) \text{ a.s.}
 \end{eqnarray*}
 Thus, we have
 \begin{eqnarray*}
&&R_H(0)= \int_{n-1} ^{\lambda + n-1}  \sigma_{t}^2(\theta)  dt  \cr
	&& =\lambda   \sum_{k=0}^\infty   \( \omega_{H1}   2 \alpha_H ^{-3}  \frac{ \alpha_H   ^{k+3} }{ (k+3) !}  -  \omega_{H2}     \alpha_H ^{-2}  \frac{ \alpha_H   ^{k+2} }{ (k+2) !} \)  \cr
	&&\quad +  \sum_{k=0}^\infty \sigma_{n-1}^2(\theta)  \lambda \left \{  \gamma_H 2   \alpha_H ^{-3}  \frac{\alpha_H ^{k+3}}{(k+3)!}  -  \alpha_H  ^{-2} \frac{\alpha_H  ^{k+2} }{ (k+2) !} + \alpha_H ^{-1}  \frac{\alpha_H ^{k+1} }{ (k+1) !} \right \}   \cr
	 && \quad +   \sum_{k=0}^\infty \frac{  \beta_H  }{    (1-\lambda)}   \lambda  \left \{   \alpha_H^{-2}  \frac{\alpha_H ^{k+2} }{ (k+2) ! }  - 2 \alpha_H^{-3} \frac{\alpha_H^{k+3} }{ (k+3)! }   \right \}   \sum_{j=1}^\infty \gamma ^{j-1}   \( \int_{n-1+\lambda-j } ^{n-j} \sigma_s  (\theta) dB_s  \)^2  \cr
	&&\quad  + \sum_{k=0}^\infty \nu_H \lambda \left \{      \alpha_H^{-2}  \frac{ \alpha_H ^{k+2} }{ (k+2) ! }  - 2  \alpha_H ^{-3} \frac{ \alpha_H ^{k+3} }{ (k+3)! }   \right \} \cr 
	&&\quad + 2  \nu_H \lambda^{-2} \sum_{k=0}^{\infty}  \int_{n-1}^{\lambda + n -1}  (\lambda^{-1}\alpha_{H} )^{k} \frac{(\lambda +n -1 -t) ^{k+2} (k+1) }{(k+2)!} Z_t ^HdW_t \cr
	&&= \lambda   \omega_{H1} 2  \varrho_{H3} - \lambda \omega_{H2} \varrho_{H2}    + \nu_H \lambda   \{    \varrho_{H2} - 2 \varrho_{H3} \}   \cr
	&& \quad + \sigma_{n-1}^2(\theta)\lambda  \{   2 \gamma_H   \varrho_{H3 }   + \varrho_{H1}  - \varrho_{H2}  \}\cr
	&& \quad  +      \lambda \frac{  \beta_H  }{    (1-\lambda)}    \left \{   \varrho_{H2}   - 2 \varrho_{H3}    \right \}   \sum_{j=1}^\infty \gamma ^{j-1}   \( \int_{n-1+\lambda-j } ^{n-j} \sigma_s  (\theta) dB_s  \)^2  \cr 
		&& \quad +2 \nu_H \alpha_H^{-2} \int_{n-1}^{\lambda + n -1}   \{  \alpha_H \lambda^{-1} ( \lambda +n-1 -t-   \lambda \alpha_H   ^{-1} ) e^{\lambda^{-1}\alpha_H (\lambda +n -1 -t) } +1 \} Z_t^H dW_t \cr
		&&=  \lambda \( \omega_H^g + \gamma h_{n-1}^H (\theta) + \alpha_H^g \lambda^{-1} \int_{n-2}^{\lambda+ n-2} \sigma_t ^2 dt + \beta_{H}^g (1-\lambda)^{-1} ( P_{n-1} - P_{\lambda+ n-2} )^2  \)  +D_n^H \text{ a.s.}
 \end{eqnarray*}
 Similarly, we  have 
 \begin{eqnarray*}
 	R_L (k) &=& \int_{\lambda+n-1} ^{n} \frac{(n-t)^k }{k!} \sigma_{t}^2(\theta)  dt  \cr
 		&=& \omega_L  (1-\lambda)^{-1} \frac{(1-\lambda) ^{k+2} }{ (k+2) !} + \sigma_{\lambda+ n-1}^2(\theta)  \left \{   \frac{\gamma_L -1 }{1-\lambda} \frac{(1-\lambda) ^{k+2} }{ (k+2) !} + \frac{(1-\lambda) ^{k+1} }{ (k+1) !} \right \}  \cr 
 		&&+ \nu_L (1-\lambda)^{-2} \( \frac{(1-\lambda)^{k+3} }{ (k+2)!} - 2 \frac{(1-\lambda)^{k+3}}{(k+3)!} \)\cr
 		&& 	 +\frac{ \alpha _L  }{(1-\lambda)^2 \lambda }   \left \{ (1- \lambda) \frac{(1-\lambda)^{k+2} }{ (k+2) ! }  - 2 \frac{(1-\lambda)^{k+3} }{ (k+3)! }   \right \}    \sum_{j=1}^\infty \gamma ^{j-1}  \int_{n-j} ^{n -j +\lambda} \sigma_s ^2 (\theta) ds  \cr
 		&& +2 \nu_L (1-\lambda)^{-2} \int_{\lambda+n-1} ^{n} \frac{(n-t)^{k+2} (k+1)}{(k+2)!} Z_t^L dW_t\cr
 			&& + 2 \beta_L (1-\lambda)^{-1}  \int_{\lambda+ n-1}^{n} \frac{(n   -t) ^{k+1}  }{(k+1)!} (P_s- P_{\lambda+ n-1} ) \sigma_s (\theta) dB_s \cr
 			&&+ \beta_L(1-\lambda)^{-1}  R_L (k+1) \text{ a.s.},
 \end{eqnarray*}
 and 
  \begin{eqnarray*}
&&R_L(0)=\int_{\lambda + n-1}^n  \sigma_{t}^2(\theta)  dt \cr
&&=  (1-\lambda) \omega_L   \varrho_{L2} + \nu_L (1-\lambda) (\varrho_{L2} -2 \varrho_{L3}) + \sigma_{\lambda+n-1}^2(\theta)  (1-\lambda)\{  (\gamma_L -1) \varrho_{L2 }   + \varrho_{L1}  \}\cr
	&& \quad   	 +(1-\lambda) \frac{ \alpha _L  }{ \lambda }   \left \{ \varrho_{L2}  - 2 \varrho_{L3}    \right \}    \sum_{j=1}^\infty \gamma ^{j-1}  \int_{n-j} ^{n -j +\lambda} \sigma_s ^2 (\theta) ds  \cr 
		&& \quad +  	2 \nu_L \beta_L^{-2} \int_{\lambda+n-1} ^{n} \{ \frac{\beta_L}{1- \lambda} ( n-t - (1-\lambda) \beta_L^{-1} ) e^{\beta_L (1-\lambda)^{-1} (n-t) } +1 \}   Z_t^L dW_t\cr
 		&&\quad + 2 \int_{\lambda+ n-1}^{n}   ( e^{\beta_L (1-\lambda)^{-1} (n-t) } -1 )(P_{t} - P_{\lambda+n-1} )  \sigma_t (\theta) dB_t \cr
		&&= (1-\lambda) \( \omega_L^g + \gamma h_{n-1}^L (\theta) + \alpha_L^g \lambda^{-1} \int_{n-2}^{\lambda+ n-2} \sigma_t ^2 (\theta) dt + \beta_{L}^g (1-\lambda)^{-1}( P_{n-1} - P_{\lambda+ n-2} )^2\) \cr
		&&\quad  +D_n^L \text{ a.s.}
 \end{eqnarray*}

 The results of $(c)$ and $(d)$ are immediate consequences of  $(a)$ and $(b)$.
  \endpf 

\textbf{Proof of Theorem  \ref{prop-integratedVol-apendix}  (b).}
To simplify the notations, we set $n=1$. 
Simple algebraic manipulations show
 \begin{eqnarray*}
 	E \[ \(D_1^H \) ^2 \middle | \FF_{0} \] &=& 4 \nu_H^2 \alpha_H^{-4} \int_{0}^{\lambda }  t \{  \lambda^{-1}\alpha_H ( \lambda -t -\alpha_H^{-1} \lambda) e^{\alpha_H \lambda^{-1} (\lambda   -t) } +1 \} ^2  dt \cr
 		&=&  \lambda^{2}   \nu_H^2 \frac{ (2\alpha_H ^2  - 8 \alpha_H  +9 ) e^{2\alpha_H  } + (16\alpha_H   - 48 ) e^{ \alpha_H } +4 \alpha_H ^2   + 22 \alpha_H   + 39   }{2 \alpha_H^6} \cr
 		&=& \lambda^2 \nu_H ^g, 
 \end{eqnarray*}
where 
\begin{eqnarray} \label{eq-nu}
	 &&  \nu_H^g= \frac{ 2 \alpha_H^6  \nu_H^2 }{     (2\alpha_H ^2  - 8 \alpha_H  +9 ) e^{2\alpha_H  } + (16\alpha_H   - 48 ) e^{ \alpha_H } +4 \alpha_H ^2   + 22 \alpha_H   + 39       } . 
\end{eqnarray}

Consider $D_{n}^{LL}$.
We first define 
\begin{eqnarray*}
	&&f_{\beta_L,1} (t) = \frac{3}{2} e^{\frac{6\beta_{L}}{1-\lambda} (1-t) } - \frac{1}{2} e^{\frac{2\beta_{L}}{1-\lambda} (1-t) } , \quad f_{\beta_L,2} (t) = \frac{1-\lambda}  {\beta_L} ( e^{\frac{\beta_L}{1-\lambda} (t-\lambda) } -1 ), \cr
	&&f_{\beta_L,3} (t) =\beta_L^{-2} [ \{ (\lambda-1) \beta_L -\lambda+1 \}  e^{\frac{\beta_L}{1-\lambda} (t-\lambda)} - \{(t-1)\beta_L  -\lambda +1\} ].
\end{eqnarray*}
By  It\^o's Lemma, we obtain 
\begin{eqnarray*}
&& E \[ \(	 \int_{\lambda } ^{1} \{ \frac{\beta_L}{1- \lambda} ( 1-t - (1-\lambda) \beta_L^{-1} ) e^{\beta_L (1-\lambda)^{-1} (1-t) } +1 \}   Z_t^L dW_t \)^2 \] \cr
&& E \[  \int_{\lambda } ^{1} \{ \frac{\beta_L}{1- \lambda} ( 1-t - (1-\lambda) \beta_L^{-1} ) e^{\beta_L (1-\lambda)^{-1} (1-t) } +1 \}^2 (t-\lambda) dt   \] \cr
&&= \frac{(1-\lambda)^2 \(  (2\beta_L^2 - 8 \beta_L + 9 ) e^{2\beta_L} + (16\beta_L-48) e^{\beta_L} + ( 4\beta_L^2 + 22 \beta_L + 39 ) \)  }{8 \beta_L^2 }
\end{eqnarray*}
and 
\begin{eqnarray*}
	  &&4E \[ \int_{\lambda}^{1}  e^{\frac{2\beta_{L}}{1-\lambda} (1-t) }  (P_{t} - P_{\lambda} )^2 \sigma_t^2 (\theta)  dt\middle | \FF_{\lambda} \] \cr
	  &&= 4 E \[   \int_{\lambda}^{1}  e^{\frac{2\beta_{L}}{1-\lambda} (1-t) }  \{ \sigma_{\lambda}^2(\theta)  + \frac{t - \lambda}{1-\lambda} (\omega_L + (\gamma_L -1) \sigma_{\lambda}^2 (\theta) ) \} \int_{\lambda}^t \sigma_s^2 (\theta) ds    dt   \middle | \FF_{\lambda} \] \cr
	 	&& \qquad + 4E \[   \int_{\lambda}^{1}  e^{\frac{2\beta_{L}}{1-\lambda} (1-t) } \frac{ \beta_L}{1-\lambda} (P_{t} - P_{\lambda} )^4  dt  \middle | \FF_{\lambda} \] \cr
	 	&&= 4E \[  \int_{\lambda}^{1}  e^{\frac{2\beta_{L}}{1-\lambda} (1-t) }  \{ \sigma_{\lambda}^2(\theta)  + \frac{t- \lambda}{1-\lambda} (\omega_L + (\gamma_L-1) \sigma_{\lambda}^2 (\theta) ) \} \int_{\lambda}^t \sigma_s^2 (\theta) ds dt   \middle | \FF_{\lambda} \]  \cr
	 	&&\qquad  + 12  E \[  \int_{\lambda}^{1} ( e^{\frac{2\beta_{L}}{1-\lambda} (1-t) } -1)    (P_{t} - P_{\lambda} )^2 \sigma_t^2 (\theta) dt  \middle | \FF_{\lambda} \] \text{ a.s.}
\end{eqnarray*}
Thus, we have
\begin{eqnarray*}
	&&E \[ \int_{\lambda}^{1}  e^{\frac{2\beta_{L} }{1-\lambda}(1-t) }   (P_{t} - P_{\lambda} )^2 \sigma_t^2 (\theta) dt   \middle | \FF_{\lambda} \] \cr
	&&= \frac{3}{2}  E \[  \int_{\lambda}^{1}     (P_{t} - P_{\lambda} )^2 \sigma_t^2 (\theta) dt  \middle | \FF_{\lambda} \] \cr
	&& \qquad- \frac{1}{2} E \[  \int_{\lambda}^{1}  e^{\frac{2\beta_{L} }{1-\lambda}(1-t) }  \{ \sigma_{\lambda}^2 (\theta) + \frac{t- \lambda}{1-\lambda} (\omega_L + (\gamma _L -1) \sigma_{\lambda}^2(\theta)  ) \} \int_{\lambda}^t \sigma_s^2 (\theta) ds dt   \middle | \FF_{\lambda} \]  \cr
	&&= E \[  \int_{\lambda}^{1}  \( \frac{3}{2} e^{\frac{6\beta_{L} }{1-\lambda}(1-t) } - \frac{1}{2} e^{\frac{2\beta_{L}}{1-\lambda} (1-t) } \)  \{ \sigma_{\lambda}^2(\theta)  + \frac{t- \lambda}{1-\lambda} (\omega_L + (\gamma_L -1) \sigma_{\lambda}^2(\theta)  ) \} \int_{\lambda}^t \sigma_s^2 (\theta) ds dt   \middle | \FF_{\lambda} \] \cr
	&&=  E \Bigg [  \int_{\lambda}^{1} f_{\beta_L,1} (t)  \{ \sigma_{\lambda}^2 (\theta) + \frac{t- \lambda}{1-\lambda} (\omega_L + (\gamma_L -1) \sigma_{\lambda}^2 (\theta) ) \}   \cr
	&& \qquad  \qquad  \times \left \{ f_{\beta_L,2} (t) \sigma_\lambda ^2 (\theta)  + f_{\beta_L,3} (t) ( \omega_L + (\gamma_L -1 ) \sigma_\lambda ^2 (\theta) ) \right \}  dt   \Bigg | \FF_{\lambda} \Bigg ] \cr
	&&= \int_{\lambda}^{1}   \(1+\frac{t- \lambda}{1-\lambda}( \gamma_L -1) \)  f_{\beta_L ,1} (t)  ( f_{\beta_L ,2} (t)    +    (\gamma_L -1) f_{\beta_L ,3} (t) )dt \sigma_\lambda^4 (\theta) \cr
	&& \quad + \int_{\lambda}^{1} \Bigg [ \(1+\frac{(t- \lambda) ( \gamma_L -1)}{1-\lambda} \)  f_{\beta_L ,1} (t)   f_{\beta_L ,3} (t)  \cr
		&&\qquad  \qquad \qquad + \frac{ (t- \lambda)  f_{\beta_L ,1} (t) }{1-\lambda} ( f_{\beta_L ,2} (t)    +     (\gamma_L -1) f_{\beta_L ,3} (t) ) \Bigg ] dt \omega_L \sigma_\lambda^2 (\theta) \cr
	&& \quad + \int_{\lambda}^{1} \frac{t- \lambda}{1-\lambda}  f_{\beta_L ,1} (t)   f_{\beta_L ,3} (t)  dt \omega_L  ^2\cr
	&&= \frac{1}{4} \( F_{\beta_L, 1} \sigma_\lambda^4 (\theta) + F_{\beta_L,2} \omega_L \sigma_\lambda^2 (\theta) + F_{\beta_L,3} \omega_L^2 \) \text{ a.s.},
\end{eqnarray*}
where the second and third equalities are due to \eqref{Lemma-var-eq01} and \eqref{Lemma-var-eq02} below, respectively, and 
\begin{eqnarray}\label{F-def}
	&&F_{\beta_L,1} =4 \int_{\lambda}^{1}   \(1+\frac{t-\lambda}{1-\lambda}( \gamma_L -1) \)  f_{\beta_L ,1} (t)  ( f_{\beta_L ,2} (t)    +    (\gamma_L -1) f_{\beta_L ,3} (t) )dt ,\cr
	&&F_{\beta_L,2} =4\int_{\lambda}^{1} \Bigg[ \(1+\frac{(t-\lambda) ( \gamma_L -1)}{1-\lambda} \)  f_{\beta_L ,1} (t)   f_{\beta_L ,3} (t)  \cr
	&&\qquad 	\qquad \qquad \qquad +  \frac{(t-\lambda)  f_{\beta_L ,1} (t)}{1-\lambda}  ( f_{\beta_L ,2} (t)    +  (\gamma_L -1) f_{\beta_L ,3} (t) ) \Bigg] dt, \cr
	&& F_{\beta_L,3} = 4\int_{\lambda}^{1} \frac{t-\lambda}{1-\lambda}  f_{\beta_L ,1} (t)   f_{\beta_L ,3} (t)  dt .
\end{eqnarray}
Hence, we arrive at 
\begin{eqnarray*}
	 G(k) &=& E \[ \int_{\lambda}^1 \frac{(1-t)^k}{k!}(P_{t} - P_{\lambda} )^2 \sigma_t^2 (\theta)   dt \middle | \FF_{\lambda} \]  \cr
	 	&=& E \[ \int_{\lambda}^1 \frac{(1-t)^k}{k!}  \{ \sigma_{\lambda}^2 (\theta) + \frac{t- \lambda}{1-\lambda} (\omega_L + (\gamma_L-1) \sigma_{\lambda}^2(\theta)  ) \} \int_{\lambda}^t \sigma_s^2 (\theta) ds   dt   \middle | \FF_{\lambda} \] \cr
	 	&&+ E \[ \int_{\lambda}^1 \frac{(1-t)^k}{k!} \frac{\beta_L}{1-\lambda} (P_{t} - P_{\lambda} )^4   \middle | \FF_{\lambda} \] \cr
	 	&=& E \[ \int_{\lambda}^1 \frac{(1-t)^k}{k!}  \{ \sigma_{\lambda}^2(\theta)  +\frac{ t- \lambda}{1-\lambda} (\omega_L + (\gamma_L -1) \sigma_{\lambda}^2(\theta)  ) \} \int_{\lambda}^t \sigma_s^2 (\theta) ds    dt  \middle | \FF_{\lambda} \] \cr
	 	&&+ \frac{6 \beta_L}{1-\lambda} E \[ \int_{\lambda}^1 \frac{(1-t)^{k+1}}{(k+1)!}  (P_{t} - P_{\lambda} )^2 \sigma_t^2 (\theta) dt   \middle | \FF_{\lambda} \] \cr
	 	&=& E \[ \int_{\lambda}^1 \frac{(1-t)^k}{k!}  \{ \sigma_{\lambda}^2(\theta)  + \frac{t- \lambda}{1-\lambda} (\omega_L + (\gamma_L -1) \sigma_{\lambda}^2(\theta)  ) \} \int_{\lambda}^t \sigma_s^2 (\theta) ds    dt  \middle | \FF_{\lambda} \] \cr
	 	&& +\frac{ 6  \beta_L}{1-\lambda} G(k+1) \text{ a.s.},
\end{eqnarray*}
where the second equality is due to It\^o's Isometric, and the third equality can be derived using arguments similar to the proofs of Theorem \ref{prop-integratedVol}.
Therefore, we obtain 
\begin{eqnarray} \label{Lemma-var-eq01}
	G(0)&=& E \[ \int_{\lambda}^1 (P_{t} - P_{\lambda} )^2 \sigma_t^2 (\theta)   dt \middle | \FF_{\lambda} \]   \cr
		&=& \sum_{k=0}^\infty  (6 \beta_L )^k E \[ \int_{\lambda}^1 \frac{(1-t)^k}{k!}  \{ \sigma_{\lambda}^2(\theta)  + \frac{t- \lambda}{1-\lambda} (\omega_L + (\gamma_L-1) \sigma_{\lambda}^2(\theta)  ) \} \int_{\lambda}^t \sigma_s^2 (\theta) ds    dt  \middle | \FF_{\lambda} \] \cr
		&=& E \[ \int_{\lambda}^1 e^{ \frac{6 \beta_L}{1-\lambda} (1-t)}  \{ \sigma_{\lambda}^2(\theta)  + \frac{t- \lambda}{1-\lambda} (\omega_L + (\gamma_L-1) \sigma_{\lambda}^2(\theta)  ) \} \int_{\lambda}^t \sigma_s^2 (\theta) ds    dt  \middle | \FF_{\lambda} \] \text{ a.s.}
\end{eqnarray}
Note that 
\begin{eqnarray*}
	 && E \[ \int_{\lambda}^t \frac{(t-s)^k }{k!} \sigma_s^2 (\theta) ds    \middle | \FF_{\lambda} \]  \cr
	 &&=  E \[ \int_{\lambda}^t  \frac{(t-s)^k }{k!} \left \{ \sigma_\lambda ^2 (\theta)  + \frac{s- \lambda}{1-\lambda}  (\omega_L + ( \gamma_L-1) \sigma_{\lambda} ^2 (\theta) ) \right \}   ds    \middle | \FF_{\lambda} \]  \cr
	 && \qquad + \frac{ \beta_L}{1-\lambda}  E \[ \int_{\lambda}^t  \frac{(t-s)^{k+1} }{(k+1)!} \sigma_s^2 (\theta) ds    \middle | \FF_{\lambda} \] \text{ a.s.},
\end{eqnarray*}
where the last equality can be derived similar to the proofs Theorem \ref{prop-integratedVol}. 
We have
\begin{eqnarray} \label{Lemma-var-eq02}
	 && E \[ \int_{\lambda}^t   \sigma_s^2 (\theta) ds    \middle | \FF_{\lambda} \]  \cr
	 &&=  E \[ \int_{\lambda}^t e ^{\frac{\beta_L}{1-\lambda} (t-s) } \left \{ \sigma_\lambda ^2 (\theta)  + \frac{s- \lambda}{1-\lambda}  (\omega_L + (\gamma_L-1) \sigma_{\lambda} ^2 (\theta) ) \right \}   ds    \middle | \FF_{\lambda} \]  \cr
	 &&=   f_{\beta_L,2} (t) \sigma_\lambda ^2 (\theta)  + f_{\beta_L,3} (t) ( \omega_L + ( \gamma_L -1) \sigma_\lambda ^2 (\theta) )  \text{ a.s.},
\end{eqnarray}
and 
\begin{eqnarray*}
 	 &&E \[ \(D_1^{LL} \) ^2 \middle | \FF_{\lambda} \] \cr
 	    &&=  F_{\beta_L, 1} \sigma_{\lambda }^4 (\theta) +  F_{\beta_L,2}  \omega_L \sigma_{\lambda }^2 (\theta) +  F_{\beta_L,3} \omega_L^2     +   (1-\lambda)^2 \nu_L^g    +  (\alpha_L^* (1-\lambda) \lambda ^{-1}   D_1^H) ^2 \cr
 	    &&  \quad +  \frac{\nu_L^2  \(  (2\beta_L^2 - 8 \beta_L + 9 ) e^{2\beta_L} + (16\beta_L-48) e^{\beta_L} + ( 4\beta_L^2 + 22 \beta_L + 39 ) \)  }{2 \beta_L^6 }   \text{ a.s.}
 	    \end{eqnarray*}
Finally, an application of the tower property leads to 
\begin{eqnarray*}
 	 &&E \[ \(D_1^{LL} \) ^2 \middle | \FF_{0} \] \cr
 	    &&=  F_{\beta_L, 1} s_{0 }^4 (\theta) +  F_{\beta_L,2}  \omega_L s_{0}^2 (\theta) +  F_{\beta_L,3} \omega_L^2   + (1-\lambda) ^2 \nu_L ^g    \text{ a.s.},
\end{eqnarray*}
where
\begin{equation}\label{eq-s}
s_0^2 (\theta) =  \omega_{H1} - \omega_{H2}  + \gamma_H \omega _L + \gamma  \sigma_{-1+ \lambda }^2 (\theta) + \beta_H h_{1}^H (\theta)   + \frac{ \gamma_H  \beta_L }{ 1-\lambda} (X_n - X_{n-1+\lambda} -\int_{n-1+\lambda}^{n} \mu_t dt )^2, 
\end{equation}
\begin{eqnarray}\label{eq-nu1}
\nu_L ^g  & =&      \frac{\nu_L^2  \(  (2\beta_L^2 - 8 \beta_L + 9 ) e^{2\beta_L} + (16\beta_L-48) e^{\beta_L}  + ( 4\beta_L^2 + 22 \beta_L + 39 ) \)  }{2 \beta_L^6 }   \cr
&&+ \{ (  \varrho_L  \beta_H (1-\lambda) \lambda ^{-1} )^2 + ( \beta_H\lambda^{-1} )^2 \}  \nu_H^g .
\end{eqnarray}
\endpf

\subsection{Proof of Theorem \ref{Thm-1}}
To easy the notations, we use $\theta$ instead of $\theta^g$ in this subsection.
 Define
  \begin{eqnarray*}
 	&&\hat{L}_{n,m} (\theta) = - \frac{1}{n} \sum_{i=1}^n   \[  \frac{ (RV_i-  \lambda \hat{h}_i  ^H(\theta ))^2 }{ \hat{\phi}_H }       +     \frac{ \( ( X_{i} - X_{\lambda+ i-1}   )^2-(1-\lambda) \hat{h}_i ^L  (\theta)  \)^2  }{\hat{\phi}_L  } \], \cr
 	 &&\hat{L}_{n } (\theta,  \phi_H, \phi_L ) =   - \frac{1}{n} \sum_{i=1}^n \[      \frac{ (IV_i-  \lambda  h_i  ^H(\theta ))^2 }{  \phi_H }   + \frac{ \( ( X_{i} - X_{\lambda+ i-1}   )^2-(1-\lambda) h_i ^L  (\theta)  \)^2  }{  \phi_L }  \],\cr
 	&& L _{n } (\theta )= -\frac{1}{n} \sum_{i=1}^n    \Bigg[ \frac{ \lambda ^2(h_i^H (\theta_0)-   h_i  ^H(\theta ))^2 +\varphi^H (\theta_0) }{ \phi_{H0} }      +  \frac{ (1-\lambda)  ^2 \( h_i ^L (\theta_0) -  h_i ^L (\theta)      \)^2 + \varphi_i ^L (\theta_0)     }{ \phi_{H0} }  \Bigg], \cr
 	 	&&\hat{s}_{n,m} (\theta ) = \frac{\partial \hat{L} _{n,m} (\theta ) }{ \partial \theta },  \,  \hat{s}_{n} (\theta,   \phi_H, \phi_L  ) = \frac{\partial \hat{L} _{n} (\theta,   \phi_H, \phi_L) }{ \partial \theta }, \,  s_{n} (\theta  ) = \frac{\partial L _{n} (\theta ) }{ \partial \theta },
 \end{eqnarray*}
 where $IV_i =   \int_{i-1}^{\lambda+ i-1} \sigma_t ^2 (\theta_0) dt$.
 Since the effect of the initial value $h_{1} (\theta )$ is of order $n^{-1}$ and thus negligible, without loss of the generality, we assume $h_1 (\theta_0 )$ is given.
 
 \begin{proposition}\label{Thm-prob}
 Under the assumption of Theorem \ref{Thm-1},  
 $\hat{\theta}  $ converges to $\theta_0$ in probability.
 \end{proposition}
 
\textbf{Proof of Proposition \ref{Thm-prob}.} 
Note that  
\begin{equation*}
	|  \hat{L}_{n,m} (\theta)  -  L _{n } (\theta)|  \leq |  \hat{L}_{n,m} (\theta)  -  \hat{L} _{n } (\theta, \theta_0)| +|  \hat{L}_{n} (\theta, \theta_0)  -  L _{n } (\theta)|. 
\end{equation*}
First consider  $|  \hat{L}_{n,m} (\theta)  -  \hat{L} _{n } (\theta, \theta_0)| $.
By Assumption \ref{Assumption1}(4), we have
\begin{eqnarray} \label{eq1-Prop1}
	E \[ \sup_{\theta } | \hat{h}_i^H (\theta )   - h_i^H (\theta) | \]  &\leq& C  \sum_{k=0}^{i-2} \gamma_u ^k  E \( |RV_{i-1-k} - IV_{i-1-k} | \)    \cr
		&\leq& C m^{-1/4},
\end{eqnarray}
and similarly, we can show
\begin{equation}\label{eq2-Prop1}
	E \[ \sup_{\theta } | \hat{h}_i^L (\theta )   - h_i^L (\theta) | \]   \leq C m^{-1/4}. 
\end{equation}
Together with Assumption \ref{Assumption1}(6), we obtain 
\begin{equation*}
	 \sup_{\theta \in \Theta} | \hat{L}_{n,m} (\theta) -  \hat{L} _{n } (\theta, \theta_0) |  = o_p (1).
\end{equation*}

Consider the second term  $|  \hat{L}_{n} (\theta, \theta_0)  -  L _{n } (\theta)|$.
We have
\begin{eqnarray*}
 &&\hat{L}_{n} (\theta, \theta_0)  -  L _{n } (\theta)  \cr
 &&=   - \frac{1}{n} \sum_{i=1}^n   \Bigg[  \frac{  \lambda D_i ^H (h_i ^H (\theta_0)  - h_i ^H (\theta) ) + (D_i^H )^2 - \varphi^H(\theta_0) }{ \phi_{H0}  }     \cr
 &&\qquad \qquad \qquad + \frac{     D_i ^{LL}    (1-\lambda)  \{ h_{i}^L (\theta_0) - h_i^L (\theta)   \} +  (D_i ^{LL}  )^2 - \varphi_i ^L (\theta_0)   }{ \phi_{L0}  } \Bigg].
\end{eqnarray*}
Note that the martingale difference terms, $D_i^H$ and  $D_i^{LL}$,  are integrable. The uniform convergence of the second term  $|  \hat{L}_{n} (\theta, \theta_0)  -  L _{n } (\theta)|$ can be established using arguments similar to the proofs of Theorem 1 in \citet{kim2016unified}.
Then, to prove the statement, we need to show the uniqueness of the maximizer of $ L _{n } (\theta)$. 
$ L _{n } (\theta)$ is concave, and the solution of the equation with its first derivative equal to zero must  satisfy $h_i ^H (\theta) = h_i ^H (\theta_0 )$ and $h_i ^L (\theta) = h_i ^L (\theta_0 )$  for all $i=1,\ldots, n$.
Thus, the maximizer $\theta^*$ must satisfy $h_i ^H (\theta^*) = h_i ^H (\theta_0 )$ and  $h_i ^L (\theta^*) = h_i ^L (\theta_0 )$  for 
all $i=1,\ldots, n$. 
Suppose that the maximizer $\theta^*$ may be different from $\theta_0$.  
Since 
\begin{eqnarray*}
	h_i ^H (\theta ) &=& \omega_H^g +  \gamma h_{i-1}^H (\theta)  +\frac{\alpha_H}{\lambda} ^g IV_{i-1} + \frac{\beta_H ^g}{ 1-\lambda } (X_{i-1} -X_{\lambda +i-2})^2 ,
\end{eqnarray*}
$\theta^*$ and $\theta_0 $ satisfy almost surely
\begin{equation*}
\bT
 \begin{pmatrix}
 \omega_{H,0}^g    - \omega_H^*   \\
 \gamma_0- \gamma^*\\ 
 \beta_{H,0}^g- \beta_H^*\\
  \alpha_{H,0} ^g  - \alpha_H^* 
 \end{pmatrix}
 =0 ,
\end{equation*}
where 
$$
\bT = 	\begin{pmatrix}
 1&  h_1 ^H(\theta_0) & (X_{1} -X_{\lambda})^2  & IV_1  \\ 
 1&  h_2 ^H(\theta_0) & (X_{2} -X_{\lambda+1})^2 & IV_2   \\ 
 \vdots& \vdots  &  \vdots &  \vdots  \\ 
 1&  h_n ^H(\theta_0) & (X_{n-1} -X_{\lambda+n-1})^2   & IV_{n-1}
 	\end{pmatrix}  .
$$
Then, since $IV_i$'s and $X_i$'s are nondegenerate random variables, we have
\begin{equation*}
 \begin{pmatrix}
 \omega_{H,0}^g    - \omega_H^*   \\
 \gamma_0- \gamma^*\\ 
 \beta_{H,0}^g- \beta_H^*\\
  \alpha_{H,0} ^g  - \alpha_H^* 
 \end{pmatrix}
 =0  \text{ a.s.},
\end{equation*}
and similarly, we obtain 
\begin{equation*}
 \begin{pmatrix}
 \omega_{L,0}^g    - \omega_L^*   \\
 \gamma_0- \gamma^*\\ 
 \beta_{L,0}^g- \beta_L^*\\
  \alpha_{L,0} ^g  - \alpha_L^* 
 \end{pmatrix}
 =0  \text{ a.s.},
\end{equation*}
which   implies  $\theta^* =\theta _0$ a.s.
This shows that the maximizer is unique. 
Finally, the result is a consequence of Theorem 1 in \citet{xiu2010quasi}.
\endpf

 \textbf{Proof of Theorem \ref{Thm-1}.} 
 The mean value theorem and Taylor expansion indicate that for some $\theta^*$ between $\theta_0 $ and $\hat{\theta} $, we have 
 \begin{equation*}
 	\hat{s}_{n,m} (\hat{\theta} ) - 	\hat{s}_{n,m} (\theta_0 ) = - 	\hat{s}_{n,m} (\theta_0 )  = -\triangledown \hat{s}_{n,m} (\theta^*) (\hat{\theta} - \theta_0 ). 
 \end{equation*} 
 By Theorem \ref{prop-integratedVol}, we have 
 \begin{equation*}	
  \begin{pmatrix}
  h_n^H (\theta_0) 
\\ 
  h_n^L (\theta_0)  
\end{pmatrix}
=   \begin{pmatrix}
  \gamma_0 + \frac{\alpha_{H,0}^g }{\lambda} & \frac{\beta_{H,0}^g}{1-\lambda}
\\ 
\frac{\alpha_{L,0} ^g}{ \lambda} & \gamma_0 + \frac{\beta_{L,0}^g}{1-\lambda}
\end{pmatrix}
  \begin{pmatrix}
  h_{n-1}^H (\theta_0) 
\\ 
  h_{n-1}^L (\theta_0)  
\end{pmatrix} 
+  \begin{pmatrix}
\omega_{H,0}^g+ \frac{\alpha_{H,0}^g}{\lambda} D_{n-1}^H + \frac{\beta_{H,0}^g}{ 1- \lambda} D_{n-1}^{LL} 
\\ 
\omega_{L,0}^g+  \frac{\alpha_{L,0}^g}{\lambda} D_{n-1}^H + \frac{\beta_{L,0}^g}{ 1- \lambda} D_{n-1}^{LL} 
\end{pmatrix}.
 \end{equation*}
Thus, by Assumption \ref{Assumption1} (1) and (5),  $h_n^H (\theta_0) $ and $h_n^L (\theta_0) $ are stationary.
Since $(X_n -X_{\lambda +n-1})^2$ and $IV_n$ are functions of $h_n^H (\theta_0) $, $h_n^L (\theta_0)$ $D_n^H$ and $D_n^{LL}$, $(X_n -X_{\lambda +n-1})^2$  and $ IV_n$ are stationary. 
 Then, similar to the proofs of Proposition \ref{Thm-prob}, we can show 
 $$
  -\triangledown \hat{s}_{n,m} (\theta^*)  \overset{p} {\to}  -\triangledown s_{n} (\theta_0).  
 $$

  Since $IV_i$'s and $X_i$'s are nondegenerate, $-\triangledown s_{n} (\theta_0)$ is almost surely positive definite. 
 By the ergodic theorem, we have 
 $$
 -\triangledown s_{n} (\theta_0) \overset{p} {\to} 2 A. 
 $$
 Thus, we obtain  
\begin{eqnarray*}
	| \hat{s}_{n,m} (\theta_0 ) - \hat{s}_{n } (\theta_0,   \phi_{H0}, \phi_{L0} )|  \leq 	| \hat{s}_{n,m} (\theta_0 ) - \hat{s}_{n } (\theta_0, \hat{\phi}_{H}, \hat{\phi}_{L} )| +  |\hat{s}_{n } (\theta_0,  \hat{\phi}_{H}, \hat{\phi}_{L} ) - \hat{s}_{n } (\theta_0,  \phi_{H0}, \phi_{L0} )|.
\end{eqnarray*}
By Assumption \ref{Assumption1} and \eqref{eq1-Prop1}--\eqref{eq2-Prop1}, we can establish  
 \begin{equation}\label{eq1-thm2}
 	| \hat{s}_{n,m} (\theta_0 ) - \hat{s}_{n } (\theta_0, ,  \hat{\phi}_{H}, \hat{\phi}_{L}  )| =  O_p(m^{-1/4} ). 
 \end{equation}
 Consider $ |\hat{s}_{n } (\theta_0,  \hat{\phi}_{H}, \hat{\phi}_{L} ) - \hat{s}_{n } (\theta_0,  \phi_{H0}, \phi_{L0} )|$.
   Simple algebraic manipulations show
\begin{eqnarray*}
 \hat{s}_{n } (\theta_0, \hat{\phi}_{H}, \hat{\phi}_{L}  ) - \hat{s}_{n } (\theta_0, \phi_{H0}, \phi_{L0} )  &=& \frac{1}{n} \sum_{i=1}^n   2 \lambda D_i ^H  \frac{\partial h_i ^H (\theta_0) }{ \partial \theta}    \(  \frac{1 }{\hat{\phi}_{H} }- \frac{1 }{\phi_{H}}      \)    \cr
  && \qquad  + 2 (1-\lambda)  D_i ^{LL} \frac{\partial h_i ^L (\theta_0) }{ \partial \theta}   \(  \frac{1 } {\hat{\phi}_{L}}  -    \frac{1 } {\phi_{L}}  \)  .
\end{eqnarray*}
Since $D_i ^H$'s are martingale differences and $h_i ^H (\theta_0)$ is $\FF_{i-1}$-adaptive, 
we have  $\frac{ 2 \lambda}{n} \sum_{i=1}^n   D_i ^H  \frac{\partial h_i ^H (\theta_0) }{ \partial \theta}    = O_p(n^{1/2})$. 
Thus, with $\|\hat{\phi}_{H} - \phi_{H0}\|_{\max} = o_p(1)$, we obtain 
\begin{equation*}
 \frac{1}{n} \sum_{i=1}^n   2 \lambda D_i ^H  \frac{\partial h_i ^H (\theta_0) }{ \partial \theta}    \(  \frac{1 }{\varphi ^H (\hat{\theta}_1)}- \frac{1 }{\varphi ^H (\theta_0)}      \) = o_p(n^{-1/2}) . 
\end{equation*}
Similarly, we can show
\begin{eqnarray*}
	&& \frac{1}{n} \sum_{i=1}^n  D_i ^{LL} \frac{\partial h_i ^L (\theta_0) }{ \partial \theta}   \(  \frac{1 } {\hat{\phi}_{L}}  -    \frac{1 } {\phi_{L}}  \)  = o_p(n^{-1/2}). 
\end{eqnarray*} 
 Hence, we have
\begin{equation} \label{eq2-thm2}
 \hat{s}_{n } (\theta_0, \hat{\phi}_{H}, \hat{\phi}_{L}  ) - \hat{s}_{n } (\theta_0, \phi_{H0}, \phi_{L0} )   = o_p (n^{-1/2}) .
\end{equation}
 By \eqref{eq1-thm2} and  \eqref{eq2-thm2}, we obtain 
 $$
 | \hat{s}_{n,m} (\theta_0 ) - \hat{s}_{n } (\theta_0,  \phi_{H0}, \phi_{L0} )|  = O_p (m^{-1/4}) + o_p (n^{-1/2}).
$$
Note that 
 \begin{eqnarray*}
  \hat{s}_{n} (\theta_0,  \phi_{H0}, \phi_{L0} ) &=&      \frac{1}{n} \sum_{i=1}^n   2 \lambda \frac{D_i ^H }{\phi_{H0}}   \frac{\partial h_i ^H (\theta_0) }{ \partial \theta}      + 2 (1-\lambda) \frac{D_i ^{LL} }{\phi_{L0}}   \frac{\partial h_i ^L (\theta_0) }{ \partial \theta}  .
 \end{eqnarray*}
 Since the martingale difference terms have the finite 4th moments, the above term convergence rate is $n^{-1/2}$.
 Thus,  \eqref{eq-result1-Thm2} is proved.
 An application of the ergodic theorem leads to  
$$
\sqrt{n}  \hat{s}_{n} (\theta_0 ) \overset{d}{\to} N (0, 4B).
$$
Finally,  we conclude 
 \begin{eqnarray*}
 	 \sqrt{n} (\hat{\theta} - \theta_0 ) &=& - \frac{1}{2}A^{-1} \sqrt{n} \hat{s}_{n} (\theta_0 )  + o_p(1) \cr
 	 	&\overset{d} {\to}& N (0, A^{-1} B A^{-1}) .
 \end{eqnarray*}
 The statement \eqref{eq-result2-Thm2} is proved. 
 \endpf


\bibliography{myReferences}
\end{spacing}